\documentstyle[preprint,tighten,prc,aps,float,epsfig]{revtex}

\begin{document}

%\draft

\title{ Probing Nucleon Strangeness with Neutrinos :\\
Nuclear Model Dependences
}
\author{
 {M.B. Barbaro}$^{a}$, {A. De Pace}$^{a}$, {T.W. Donnelly}$^b$ 
 {A. Molinari}$^{a}$, and {M.J. Musolf}$^{c,d}$
}
\address{
 $^a$ Dipartimento di Fisica Teorica dell'Universit\`a 
 di Torino and \\
 Istituto Nazionale di Fisica Nucleare, Sezione di Torino, \\
 via P.Giuria 1, I-10125 Torino, Italy \\
 $^b$ Center for Theoretical Physics, \\
 Laboratory for Nuclear Science and Department of Physics, \\
 Massachusetts Institute of Technology, Cambridge, MA 02139, USA\\
 $^c$ Continuous Electron Beam Accelerator Facility Theory Group, MS 12H2,
 Newport News, Virginia 23606, USA \\
 $^d$ Institute for Nuclear Theory, University of Washington, Seattle, WA 98195,
 USA
}
%\date{May 1996}

\maketitle

\vskip -2mm
\begin{abstract}

\vskip -4mm
	The extraction of the nucleon's strangeness axial charge, $\Delta s$,	
from inclusive, quasielastic neutral current neutrino cross sections
is studied within the framework of the plane-wave impulse approximation. We 
find that the value of $\Delta s$ can depend significantly on the choice of
nuclear model used in analyzing the quasielastic cross section. This 
model-dependence may be reduced by one order of magnitude when $\Delta s$
is extracted from the ratio of total proton to neutron yields. We apply this
analysis to the interpretation of low-energy neutrino cross sections and
arrive at a nuclear theory uncertainty of $\pm 0.03$ on the value of $\Delta s$
expected to be determined from the ratio of proton and neutron yields measured
by the LSND collaboration. This error compares favorably with estimates of
the SU(3)-breaking uncertainty in the value of $\Delta s$ extracted from
inclusive, polarized deep-inelastic structure function measurements. We also
point out several general features of the quasielastic neutral current neutrino
cross section and compare them with the analogous features in inclusive,
quasielastic electron scattering. 
\end{abstract}

%\pacs{25.30-c, 25.30.Pt}

\vfill
\begin{center}
submitted to {\em Physical Review C}
\end{center}
\vfill
CTP\#2507, \ \ INT DOE/ER/40561-243-INT96-00-112 \hfill May 1996

\eject

\vglue 1cm
\vfill 
\noindent
{\small $^*$This work is supported in part by
funds provided by the U.S. Department of Energy (D.O.E.) under contracts
\#DE-AC05-84ER40150, \#DE-FG06-90ER40561, CEBAF and INT grants, and
cooperative agreement \#DE-FC02-94ER40818. M.J.M. was also supported by the
National Science Foundation National Young Investigator Program.}

\noindent
{\small $^\dagger$On leave from
 the Department of Physics, Old Dominion University, Norfolk, VA 23529 USA} 
\eject

\section{Introduction}
\label{sec:intro}

Inclusive, quasielastic (QE) weak neutral-current (NC) reactions have received 
considerable attention recently as a means of probing the strange quark content
of the nucleon\cite{Mus94}.
In particular, analyses of the Brookhaven experiment 
E734\cite{Ahr87,Gar93a,Hor93} have generated bounds on the nucleon's 
strangeness axial-vector matrix element which are essentially consistent 
\cite{MAD}
with the values for $\Delta s$, the strange-quark
contribution to the nucleon spin determined from polarized deep inelastic
scattering\cite{Ash89,Ade93,Ant93,Ada94,Abe95}. 
Less stringent bounds on the strangeness vector current form factors have
also been extracted from the Brookhaven data\cite{Gar93a,Hor93}. 
In a similar vein, one
expects a determination of the ratio of proton and neutron yields in the
LSND experiment at Los Alamos\cite{Lou} 
to produce even more stringent limits on
some of these form factors\cite{Gar92,Gar93b}. 
The results from these studies should complement
the results from several parity-violating (PV) elastic and QE electron
scattering measurements currently underway at MIT-Bates
\cite{Sam,Pitt} and planned for both CEBAF\cite{Bei,Bec,Fin} and 
Mainz\cite{Mai}.
Indeed, this program of semileptonic NC scattering measurements affords one
with a unique, low-energy window on the non-valence quark structure of the
nucleon\cite{Mus94}. 

Since the neutrino NC reactions of interest require the detection of a
final-state nucleon knocked-out of a nuclear target, a proper interpretation
of the results in terms of single-nucleon physics requires that one have a 
sufficiently reliable understanding of the nuclear, many-body impact
on the neutrino cross sections. 
Thus far, nuclear calculations have been performed using a relativistic Fermi
Gas (RFG) model\cite{Hor93} and a continuum RPA approach\cite{Gar93b}.
In the present paper,
we place these studies in the context of a more general framework by pointing
out several features of QE neutral current scattering not previously realized in
the literature.  
Specifically, we note the complementarity of inclusive, QE NC electron
scattering, in which the outgoing electron is detected, and QE NC neutrino
scattering, in which the detected particle is a nucleon. 
These two processes -- to which we refer as $t$-inclusive and $u$-inclusive 
scattering, respectively -- explore 
different regions of the two-parameter missing energy (${\cal E}$) and missing
momentum ($p$) space.
Consequently, in the plane-wave impulse approximation (PWIA), these two types 
of QE NC
scattering may display different sensitivities to the many-body physics which
enters the one-body spectral function, $S(p,{\cal E})$. 

We use this framework to arrive at the first (to our knowledge)
published estimate of the nuclear theory uncertainty 
associated with the extraction of $\Delta s$ from the BNL and LSND measurements.
Our approach in doing so is the following. By varying the nuclear model
used in analyzing the QE cross section, we change $S(p,{\cal E})$ and, 
consequently, obtain different 
extracted values for $\Delta s$. To be as conservative as possible, we
choose two simple, tractable models lying near the extremes of the 
spectrum of reasonable nuclear models. Specifically, we employ the RFG model 
and a hybrid model (HM) involving harmonic oscillator shell model wave 
functions for the bound nucleons and plane waves for the continuum states. 
The former represents the ``maximally unconfined'' extreme, since it 
employs plane waves for bound and continuum single-particle states; 
additionally, as discussed below, only the on-shell electroweak current 
matrix elements occur\cite{Cen96}.  The latter is ``overconfined'' in the 
sense that a harmonic oscillator basis is used, implying that the long-range 
behavior of the bound single-particle wave functions is gaussian rather 
than exponential as should be the case with finite potentials. Our approach 
in the present work is to model only these extremes to explore the 
``worst-case scenario'' for extracting $\Delta s$ from QE neutrino scattering. 

In fact, these two models reproduce rather
well the experimental QE response for inclusive electron 
scattering, especially for integrated quantities such as the Coulomb sum 
rule\cite{Jou95} or the responses discussed in the present work, 
despite the significant differences found in the behavior of the respective 
spectral functions in $({\cal E}, p)$ space. We expect that the spread in
extracted values of $\Delta s$ obtained using more sophisticated
treatments of the response, such as those which include the effects of
correlations and more realistic single-particle wave functions, will be
reasonably characterized by 
the difference between the RFG and HM values. We take as the 
nuclear theory error, $\delta_{\hbox{nuc}}(\Delta s)$, the difference
between $\Delta s$(RFG) and $\Delta s$(HM). We find that at the kinematics 
of the LSND experiment, $\delta_{\hbox{nuc}}(\Delta s)\approx \pm 0.25$
when the individual proton or neutron knockout cross sections are
used. This error is
roughly as large in magnitude as the average value for $\Delta s$
determined from the deep-inelastic measurements. If, however, one considers
the ratio $R_\nu$
of proton to neutron yields rather than the individual cross 
sections, as has been proposed for the interpretation of the LSND data
\cite{Gar92,Gar93b}, 
we find that the magnitude of $\delta_{\hbox{nuc}}(\Delta s)$ is reduced by
more than an order of magnitude. In this case, the nuclear theory uncertainty is
significantly smaller than estimates of the theoretical SU(3)-breaking
uncertainty in the deep inelastic values for $\Delta s$ \cite{Jen91,Dai95}.
{}From this standpoint, our results complement those of Ref. \cite{Gar93b}
which analyzed the impact of final-state interactions (FSI) on the extraction
of $\Delta s$ from QE neutrino cross sections and which found that use of
$R_\nu$, as compared with the individual cross
sections, significantly reduces one's sensitivity to distortion in the 
outgoing nucleon's wave function.

In the remainder of this paper, we discuss these features in
more detail. A reader primarily interested in the application of our
analysis to the extraction of $\Delta s$ is encouraged to read Sections
\ref{sec:nuclmodels} and \ref{sec:axialff},
along with Eqs.~(\ref{eq:dsigma6}-\ref{eq:d2sigmau}) of Sec.~\ref{sec:IA}. 
The reader
interested as well in the general features of $u$- and $t$-channel QE
cross sections is directed to Sec.~\ref{sec:tu}, where 
we discuss the difference between QE neutral current
neutrino and electron scattering in general terms; Sec.~\ref{sec:IA}, in which
we consider the implications of this difference for the (PWIA) analysis of QE
scattering; and Sec.~\ref{sec:RFG}, where we specify further to the RFG.
Section \ref{sec:concl} summarizes our conclusions and is followed by an 
Appendix.

\section{ Inclusive $\lowercase{\bbox{t}}$- and 
$\lowercase{\bbox{u}}$-channel scattering}
\label{sec:tu}

The leading order exclusive QE cross section is generated by the Feynman
amplitude associated with the diagram of Fig.~\ref{fig:Fig1}. 
Here, a lepton $\ell$ scatters off an $A$-body nucleus to a final lepton state
$\ell'$ via the exchange of a vector boson $V$. 
In the scattering, a nucleon $N$ is knocked out leaving behind an $(A-1)$-body
daughter nucleus generally in an excited state. 
We let $K^\mu=(\epsilon, \bbox{k})$ and 
$K^{\prime\ \mu}=(\epsilon', {\bbox{k}}^{\prime})$ denote the initial and final 
lepton momenta, respectively, 
$P_N^\mu=(E_N, \bbox{p}_N)$ the four-momentum of the ejected 
nucleon and $-\bbox{p}$ the three-momentum of the recoiling daughter nucleus. 

Following\cite{Day} we define the missing energy ${\cal E}$ as 
\begin{equation}
  {\cal E}\equiv\sqrt{{\bbox{p}}^{2}+{M_{A-1}^{\ast 2}}}
    -\sqrt{{\bbox{p}}^{2}+{M_{A-1}^2}}\ ,
\label{eq:men}
\end{equation}
where ${M_{A-1}}$ and ${M_{A-1}^\ast}$ are the masses of the recoiling nucleus 
in its ground and excited states respectively. 
Thus the missing energy used in this work actually corresponds to the 
excitation energy of the
residual nucleus in a frame where it is moving with a momentum $-\bbox{p}$.
The conditions for four-momentum conservation in the laboratory frame (target
nucleus at rest) read
\begin{mathletters}%
\begin{eqnarray}
  \epsilon+M_A &=& \epsilon'+E_N+{\cal E}+\sqrt{{\bbox{p}}^{2}+{M_{A-1}^2}} \\
\label{eq:conse}
  \bbox{k} &=& {\bbox{k}}^{\prime}+\bbox{p}_N-\bbox{p}\ .
\label{eq:consp}
\end{eqnarray}
\end{mathletters}%
If we further introduce the nucleon kinetic energy $T_N=E_N-m_N$, the nuclear 
recoil energy ${T_{A-1}}=\sqrt{{\bbox{p}}^{2}+{M_{A-1}^2}}-{M_{A-1}}$ and the 
positive nuclear separation energy $E_S={M_{A-1}}+m_N-M_A$, then one may rewrite
Eq.~(\ref{eq:conse}) as
\begin{equation}
  \epsilon=\epsilon'+T_N+{\cal E}+{T_{A-1}}+E_S\ .
\label{eq:eps}
\end{equation}

At this point, we refer to Eqs.~(\ref{eq:consp}) and (\ref{eq:eps}) along with 
Fig.~\ref{fig:Fig1} to describe the difference between QE $(e,e')N$ and 
QE $(\nu, N)\nu'$ kinematics.  
In the former case, the initial and final lepton energies and three-momenta
are fixed experimentally. 
Thus, one may specify a given value of the energy transfer 
$\omega=\epsilon-\epsilon'$ and three-momentum transfer 
$\bbox{q}=\bbox{k}-{\bbox{k}}^{\prime}$ by properly selecting the beam energy 
and momentum and the lepton detector settings. 
The experimentally-fixed Lorentz invariant for this process is
$Q^2=\omega^2-{\bbox{q}}^{2}<0$ and we correspondingly refer to this process as
``$t$-inclusive'' scattering. 
Further, by combining Eqs.~(\ref{eq:consp}) and (\ref{eq:eps}), we may obtain
the following relation between the missing energy and recoil
momentum: 
\begin{equation}
  {\cal E}(\bbox{p}) = 
    \omega-{T_{A-1}}-E_S-\left[\sqrt{m^2_N+(\bbox{p}+\bbox{q})^2}-m_N\right] \ ,
\label{eq:men2}
\end{equation}
where the quantity in the square brackets is just $T_N$. 
Note that ${T_{A-1}}$ also carries a dependence on $p^2$. 
However, for typical targets in QE scattering ({\em e.~g.},$ ^{12}$C), 
${T_{A-1}}$ is much smaller than the other energies involved in the problem so 
that one may usually neglect it without introducing any significant error.
Eq.~(\ref{eq:men2}) actually defines a continuous family of curves,
parameterized by the angle between $\bbox{p}$ and $\bbox{q}$. 
This family is accordingly bounded by the curve we denote by ${\cal E}^{-}$,
corresponding to $\cos(\hat p\cdot\hat q)=-1$ ($\bbox{p}$ and $\bbox{q}$ 
anti-parallel) and the curve ${\cal E}^{+}$, for which 
$\cos(\hat p\cdot\hat q)=1$ ($\bbox{p}$ and $\bbox{q}$ parallel).

To get the $t$-inclusive cross section, one must integrate over the
three-momenta of the undetected particles (daughter nucleus and 
outgoing nucleon). As we show below, 
this corresponds to integrating over the region in the $({\cal E}, p)$ space 
lying between the two curves ${\cal E}^{+}$ and ${\cal E}^{-}$.
In fact two situations may actually occur, depending upon the sign of the 
vertical intercept of these curves: 
\begin{eqnarray}
 I_t&\equiv& {\cal E}^+(0) = {\cal E}^-(0) 
\nonumber
\\
& = & \omega - T_{A-1} -E_S - (\sqrt{q^2+m_N^2}-m_N) \ .
\label{eq:eps0}
\end{eqnarray}
For $I_t \leq 0$, the integration region, which we denote ${\cal D}$, will be 
bounded by ${\cal E}^-(p)$ and the $p$-axis (Fig.~\ref{fig:Fig2}a), while,
for $I_t\geq 0$, ${\cal D}$ is bounded by ${\cal E}^-(p)$, ${\cal E}^+(p)$ and 
the $p$-axis (Fig.~\ref{fig:Fig2}b).
Note that in the figures, as an orientation, we have taken typical values for
$q$ (500 MeV/c), $\omega$ (100 and 170 MeV, respectively) and $E_S=8$ MeV.
Furthermore, we have included only the portions of the curves existing in the 
upper right quadrant, since both ${\cal E}$ and $p$ are by definition 
non-negative. For this region, the maximum value of the 
missing energy occurs for \cite{Cen96}
\begin{equation}
  p = p_{max} = q \frac{{M_{A-1}}}{{M_{A-1}}+m_N}
\label{eq:pmax}
\end{equation}
and has the value 
\begin{eqnarray}
& {\cal E}&(p_{max}) \equiv {\cal E}_{max}
\nonumber
\\
& = & \omega-E_S - \left[\sqrt{q^2+ \left({M_{A-1}}+m_N\right)^2} -
\left({M_{A-1}}+m_N\right)\right]
\nonumber
\\
& \simeq & \omega-E_S \ . 
\label{eq:emax}
\end{eqnarray}
For typical QE kinematics, one has ${\cal E}_{max}\approx 100$ MeV. 

Parenthetically, we note that the region ${\cal D}$ depends on
the independent kinematic variables for the inclusive process. For
example, when the cross section for an outgoing electron of a specified
energy and scattering angle is considered, one has ${\cal D}={\cal D}(
\epsilon', \theta_e)$. When the energy- or angle-integrated cross section
is of interest, the region ${\cal D}$ consists of the union of all regions
${\cal D}(\epsilon', \theta_e)$ allowed by the kinematics.

In the case of QE NC neutrino scattering, the final lepton is undetected
and the four-momentum of the outgoing nucleon is specified \footnote[1]{
In practice, detectors include the full solid angle for the outgoing nucleon.}.
Thus, the experimentally-fixed variables become
\begin{mathletters}%
\begin{eqnarray}
  u_0 &=& \epsilon-E_N \\
\label{eq:u0}
  \bbox{u} &=& \bbox{k}-\bbox{p}_N\ .
\label{eq:uv}
\end{eqnarray}
\end{mathletters}%
The corresponding Lorentz invariant in this case is $U^2=u_0^2-|\bbox{u}|^2$
and we refer to this process as `` $u$-inclusive'' scattering. 
{}From the conservation relations (Eqs.~(\ref{eq:conse}) and (\ref{eq:consp})), 
one has
\begin{mathletters}%
\begin{eqnarray}
  u_0 &=& {\cal E}+\epsilon'+{T_{A-1}}+E_S-m_N \\
\label{eq:u02}
  \bbox{u} &=& {\bbox{k}}^{\prime}-\bbox{p}\ .
\label{eq:uv2}
\end{eqnarray}
\end{mathletters}%
Since the neutrino is massless, one has that 
$\epsilon'=|{\bbox{k}}^{\prime}|=|\bbox{u}+\bbox{p}|$, so that 
Eq.~(\ref{eq:uv2}) may be re-written as
\begin{equation}
  {\cal E}(p)=u_0-E_S+m_N-{T_{A-1}}-|\bbox{u}+\bbox{p}|\ .
\label{eq:men3}
\end{equation}
Eq.~(\ref{eq:men3}) is the $u$-channel analog of the relation in 
Eq.~(\ref{eq:men2}).
Neglecting the small $p^2$-dependent recoil kinetic energy and defining a
quantity $\delta$ as
\begin{equation}
  \delta\equiv u_0-E_S+m_N\ ,
\label{eq:delta}
\end{equation}
we obtain the following boundaries for the region ${\cal D}$ over which one 
should integrate to get the $u$-inclusive cross section
\begin{mathletters}%
\begin{eqnarray}
  {\cal E}^+ &=& \delta-(u+p),\ \ \ \ \mbox{$\bbox{u}$ and $\bbox{p}$ parallel}
    \\
\label{eq:men4}
  {\cal E}^- &=& \delta-|u-p|=
    \left\{
    \begin{array}{cl}
  {\cal E}_>^- =\delta-(u-p) 
      \  \ \ &\mbox{$\bbox{u}$ and $\bbox{p}$ anti-parallel, $u>p$}\\
  {\cal E}_<^- =\delta-(p-u) 
      \  \ \ &\mbox{$\bbox{u}$ and $\bbox{p}$ anti-parallel, $u<p$}
    \end{array}
    \right.
\label{eq:men5}
\end{eqnarray}
\end{mathletters}%
We note that the reason for the appearance of two cases for ${\cal E}^-$
is that the undetected particle in this instance is massless. Consequently,
$\epsilon'$ -- and therefore ${\cal E}$ -- depends linearly on 
$|\bbox{u}+\bbox{p}|$. In the $t$-channel case, the un-detected outgoing
particle is massive, rendering the dependence of ${\cal E}$ on
$|\bbox{q}+\bbox{p}|$ quadratic and the relative magnitudes of
$q$ and $p$ inconsequential.

To assess the extent and the shape of the region ${\cal D}$ it helps to
observe that in the $u$-inclusive scattering the maximum of the curve ${\cal
E}^{-}$  occurs for
\begin{equation}
  p = p_{max} =u
\label{eq:pmaxu}
\end{equation}
where ${\cal E}^{-}(p)$ assumes the value
\begin{equation}
  {\cal E}^{-}(p_{max}) = \delta \ .
\label{eq:eplusmax}
\end{equation}
The relations in Eqs.~(\ref{eq:pmaxu}) and (\ref{eq:eplusmax}) are the
$u$-channel versions of Eqs.~(\ref{eq:pmax}) and (\ref{eq:emax}). In
both cases, the location of the maximum depends linearly on the magnitude
of the independent vector quantity ($q$ or $u$) while the height of the
maximum is linear in the independent time-like quantity ($\omega$ or
$u_0$). 

For future reference, we specify 
the two values of the momentum ($p_<^-$ and $p_>^-$)
where ${\cal E}^{-}$ vanishes.
They are given by
\begin{equation}
  p_<^- = u-\delta \ \ \ \ \ \mbox{and}\ \ \ \ \ p_>^- = u+\delta\ ,
\label{eq:pminus}
\end{equation} 
$p_>^-$ being positive definite whereas $p_<^-$ can be either positive or
negative. On the other hand ${\cal E}^+(p)$ vanishes for
\begin{equation}
  p^+ = \delta-u\ ,
\label{eq:pplus}
\end{equation} 
which can be positive or negative.

Again in complete analogy with the $t$-channel
the curves ${\cal E}^{-}(p)$ and ${\cal E}^{+}(p)$ intercept each other at 
$p=0$, where their common value is
\begin{equation} 
  I_u\equiv{\cal E}^{-}(0)={\cal E}^{+}(0)=\delta-u\ .
\label{eq:epsmp0}
\end{equation}
Let us denote by ${\cal D}(\theta_N)$
the allowed region in the $({\cal E}, p)$ plane at fixed 
$\theta_N$, the angle between $\bbox{k}$ and $\bbox{p}_N$.\footnote[2]{As in
the $t$-channel case, the integration region depends on the two independent
kinematic variables. In the present context, it is useful to choose one of
them to be $\theta_N$ and to suppress the dependence of ${\cal D}$ on the
other.} 
Two situations can then occur corresponding to two different shapes for this 
region: Either $I_u$ is positive or is negative. 
In the latter case $p_<^-$ is positive and ${\cal D}(\theta_N)$ is a triangle, 
lying of course in the physical quadrant of the $({\cal E}, p)$ plane.
The area of the triangle is $\delta^2$; hence, it is fixed once the moduli of 
the momenta of the incoming neutrino and outgoing nucleon are given.
On the other hand the {\em position} of the triangle in the $({\cal E},p)$ plane
depends on $\theta_N$, as well as on $k=|\bbox{k}|$ and 
$p_N=|\bbox{p}_N|$ (indeed, according to Eq.~(\ref{eq:pmaxu}), the upper 
vertex of the triangle is fixed by $u$, whose value also depends on these 
variables (see Eq.~(\ref{eq:uv}))).

This situation is illustrated in Fig.~\ref{fig:Fig3}a for kinematics
typical of the LSND experiment, namely by choosing $\epsilon$ = 200 MeV and 
$T_N=$ 60 MeV ($p_N =$ 341 MeV/c).
We thus see that when $\bbox{p}_N$ and $\bbox{k}$ are antiparallel
($\theta_N=180^\circ$), $I_u$ is 
indeed negative and ${\cal D}$ is given by the triangle on the right-hand side
of the figure. 

As $\theta_N$ is varied from antiparallel ($180^\circ$) to parallel ($0^\circ$) 
while keeping $k$ and $p_N$ fixed the value of $u$ decreases 
and the triangle moves continuously in the leftward direction.  
When $I_u$ is positive, then ${\cal D}(\theta_N)$ becomes quadrangular, as 
displayed in Fig.~\ref{fig:Fig3}b for the case in which $\bbox{k}$ 
and $\bbox{p}_N$ are parallel and for kinematics typical of the BNL experiment.
The area of the quadrangle is given by $2u\delta-u^2$.

For an experimental situation in which nucleons are detected over the full 
$4\pi$ solid angle (as in LSND), the global integration region ${\cal D}$ will 
be given by taking the union of sub-regions ${\cal D}(\theta_N)$ for all 
$\theta_N$. For example, as illustrated in the two panels of 
Fig.~\ref{fig:Fig3} for cases with $I_u<0$ and $I_u>0$, respectively, the 
regions ${\cal D}$ are defined by the lines whose vertices are labeled 
$ABCD$ ($I_u<0$: quadrangle) and $ABCDE$ ($I_u>0$: pentagon), respectively.

\section{ Plane-Wave Impulse Approximation} 
\label{sec:IA}

Much of the interest in the inclusive $(e, e')N$ and 
$(\nu, N)\nu'$ scattering reactions stems from an interpretation of the
cross sections employing the PWIA. Invoking the PWIA
corresponds to modeling the shaded vertex in Fig.~\ref{fig:Fig1} as 
illustrated in Fig.~\ref{fig:Fig4}.
Here, one assumes that only one nucleon participates in the
scattering by absorbing the virtual vector boson $V$, the remainder of the
nucleus acting just as a spectator. 
Three-momentum conservation in the laboratory system requires that the struck
nucleon has momentum $\bbox{p}$ since the initial nucleus is at rest and the
daughter nucleus recoils with momentum $\bbox{p}_{A-1}=-\bbox{p}$. 
Before absorbing the $V$, the struck nucleon has energy $E$ and
in general does not lie on the mass-shell. 
In the limit that final-state interactions are neglected, as is indeed the
case in the PWIA, the outgoing nucleon is assumed to be on the mass-shell with
an energy $E_N=\sqrt{\bbox{p}_N^2+m^2_N}$. 
Under these assumptions, the differential cross section can be written in terms
of kinematic and phase space factors, the square of the invariant amplitude
for scattering of the incident lepton from a single nucleon --- 
half-off-shell, since the struck nucleon is in general 
off-shell\cite{Deforest} --- and a function
$S(p,{\cal E})$, referred to as the spectral function, which carries 
information on
the probability of finding a nucleon inside the nucleus with momentum 
$\bbox{p}$ and energy 
\begin{equation}
  E = M_A-\sqrt{{\bbox{p}}^{2}+{M_{A-1}^2}}-{\cal E}\ ,
\end{equation}
It reads:
\begin{eqnarray}
  d\sigma & = & \frac{1}{4kM_A}\overline{|{\cal M}|}^2
   \frac{d^3k'}{(2\pi)^3 2\epsilon'}\frac{d^3 p_N}{(2\pi)^3 2E_N}
   \frac{d^3p_{A-1}}{(2\pi)^3 2 E_{A-1}^\ast}
\nonumber
\\
& \times & (2\pi)^4 \delta^{(4)}(K+P_A-K'-P_N-P_{A-1})
\label{eq:dsigma1}
\end{eqnarray}      
where $P^\mu_A=(M_A, \bbox{0})$ is the four-momentum of the
target nucleus, $P^\mu_{A-1}=(E_{A-1},-\bbox{p})$ is the four-momentum
of the daughter nucleus, ${\cal M}$ is the invariant lepton-nucleus scattering
amplitude, and where the bar over ${\cal M}$ denotes the appropriate average
over initial spins and sum over final spins. Performing the integral over
$\bbox{p}_{A-1}$
gives
\begin{eqnarray}
  d\sigma & = & \frac{1}{4kM_A}\overline{|{\cal M}|}^2
    \frac{d^3k'}{(2\pi)^3 2\epsilon'}\frac{d^3 p_N}{(2\pi)^3 2E_N}
    \frac{2\pi}{2 E_{A-1}^\ast}
\nonumber
\\
& \times & \delta(\epsilon+M_A-\epsilon' -E_N-E_{A-1}^\ast)\ .
\label{eq:dsigma2}
\end{eqnarray}      
The general form for the square of the invariant amplitude is
\begin{equation}
  \overline{|{\cal M}|}^2=g^4 D_V(Q^2)^2 L_{\mu\nu} W^{\mu\nu}\ \ \ ,
\label{eq:msq}
\end{equation}
where $g$ is the strength of the fermion-vector boson coupling, 
$D_V(Q^2)=(Q^2-M_V^2+i\varepsilon)^{-1}$, $M_V$ is the electroweak vector boson
mass, $L_{\mu\nu}$ is the usual leptonic tensor appearing in semi-leptonic
scattering and $W_{\mu\nu}$ is the corresponding hadronic tensor. 
They are given by the following well-known expressions:
\begin{mathletters}%
\begin{eqnarray}
  L_{\mu\nu}&=&\frac{1}{8}{\rm Tr}\left[{\bar u}(k',s')\Gamma_\mu
    u(k,s){\bar u}(k',s')\Gamma_\nu u(k,s)\right]\\
\label{eq:lmunu}
  W_{\mu\nu}&=&\langle A| {\hat J}_\mu^{\dag}|f\rangle\langle f| 
    {\hat J}_\nu|A\rangle\ ,
\label{eq:wmunu}
\end{eqnarray}
\end{mathletters}%
where $|A\rangle$ is the initial target nucleus, $|f\rangle$ is the final
hadronic state, $\Gamma_\mu$ is a Lorentz structure associated with the 
lepton-vector boson vertex,
${\hat J}_\lambda$ is the hadronic current operator, and
where the appropriate average over initial and sum over final hadronic
states is understood in Eqs.~(\ref{eq:dsigma2})-(\ref{eq:wmunu}).
Furthermore, in Eq.~(\ref{eq:lmunu}) the spinors are normalized according to the
Bjorken and Drell conventions\cite{Bjo64} and the weak vertices are given by 
the usual expressions.

In the PWIA framework only the one-body component of the nucleonic current, 
namely
\begin{eqnarray}
  {\hat J}_\lambda &=&\sum_{s,s'}
    \int\frac{d^3k}{(2\pi)^3}\int\frac{d^3k'}{(2\pi)^3}
    \frac{1}{2{E_k}}\frac{1}{2{E_{k'}}} 
    (2\pi)^3\delta({\bbox{k}}^{\prime}-\bbox{k}-\bbox{q})
    \nonumber \\
  &&\times{\bar u}_N(k', s') \Gamma_\lambda^N u(k,s){\hat a}^{\dag}(k',s')
    {\hat a}(k,s) \ ,
\label{eq:jmu}
\end{eqnarray}
is kept.
In the above ${\hat a}^{\dag}(k,s)$ and ${\hat a}(k,s)$ are, respectively, the
operators that create and annihilate bound nucleons ({\em i.e.,} in momentum 
space, having Fourier component $k$ and spin projection $s$) 
and $\Gamma_\lambda^N$ is the Lorentz structure associated with the 
single-nucleon current matrix element\cite{Deforest}. 
Normalizing the states according to Ref.~\cite{Bjo64},
\begin{equation}
  \langle\bbox{p}|\bbox{q}\rangle = (2\pi)^3 2 E_{\bbox{p}}\ 
     \delta(\bbox{p}-\bbox{u}) \ ,
\label{eq:pq}
\end{equation}
and substituting the expression in Eq.~(\ref{eq:jmu}) into Eq.~(\ref{eq:wmunu})
yields 
\begin{eqnarray}
  W_{\mu\nu} & = &
    \sum_{s,s'} \left(\frac{1}{2{E_p}}\right)^2\langle A| {\hat a}^{\dag}(p,s)
    |A-1\rangle\langle A-1|{\hat a}(p,s)|A\rangle 
\nonumber
\\
& \times & w_{\mu\nu}(\bbox{p},\bbox{p}_N) \ ,
\label{eq:wbig}
\end{eqnarray}
where
\begin{equation}
  w_{\mu\nu}(\bbox{p},\bbox{p}_N) = 
    {\bar u}(p,s)\Gamma_\mu u(p_N,s'){\bar u}(p_N,s')\Gamma_\nu u(p,s)
\label{wsmall}
\end{equation}
is the so-called single-nucleon tensor and the ket $|A-1\rangle$ represents the
daughter nucleus in either its ground state or one of its excited states. 
The replacement of $E_{A-1}$ with ${\cal H}\equiv{\hat H}-E_{A-1}^{(0)}$ inside 
the $\delta$-function appearing in Eq.~(\ref{eq:dsigma2}) allows one to 
perform a sum over all daughter-nucleus states by making use of the closure 
relation. One then obtains
\begin{eqnarray}
  d\sigma &=& \frac{1}{4k M_{A}}(2\pi) \left(\frac{1}{2E}\right)^2
    g^4 D_V(Q^2)^2 L_{\mu\nu}\sum_{s,s'} w^{\mu\nu}(\bbox{p},\bbox{p}_N)
\nonumber
\\
&& \!\!\!\! \!\!\!\! \!\!\!\! \!\!\!\! \times
  \langle A|{\hat a}^{\dag}(p,s)\delta(\epsilon+M_{A}-\epsilon'
    -E_N-E_{A-1}^0-\hat{\cal H}){\hat a}(p,s)|A\rangle 
    \frac{d^3k'}{(2\pi)^3 2\epsilon'}\frac{d^3 p_N}{(2\pi)^3 2E_N} \ .
\nonumber\\
\label{eq:dsigma3}
\end{eqnarray}
Noticing that $E=E_N-\epsilon+\epsilon'$ and defining the chemical potential 
$\mu\equiv M_A-E_{A-1}^0$ (not to be confused with a Lorentz index),
one has 
\begin{eqnarray}
  d\sigma & = & \frac{(2\pi)^4}{2k}\left(\frac{1}{2E}\right) g^4 D_V(Q^2)^2 
    L_{\mu\nu} w^{\mu\nu}(\bbox{p},\bbox{p}_N) 
\nonumber 
\\
& \times & S(p,{\cal E})\frac{d^3k'}{(2\pi)^3 2\epsilon'}
\frac{d^3 p_N}{(2\pi)^3 2E_N}\ ,
\label{eq:dsigma5}
\end{eqnarray}
where the spectral function is
\begin{eqnarray}
  S(p,{\cal E}) & = & \frac{1}{(2\pi)^3}\left(\frac{1}{2E}\right)
    \left(\frac{1}{2 M_A}\right)
\nonumber
\\
& \times & \langle A|{\hat a}^{\dag}(p,s)\delta(E+\hat{\cal H}-\mu)
{\hat a}(p,s) |A\rangle \ .
\label{eq:S}
\end{eqnarray}
A further elementary elaboration of Eq.~(\ref{eq:dsigma5}) leads to the 
exclusive cross section
\begin{eqnarray}
  \frac{d^4\sigma}{d\Omega' dk' d\Omega_N dE_N} & = &
    \frac{1}{(2\pi)^2} \frac{1}{2k} \frac{1}{2E} g^4 D_V(Q^2)^2
\nonumber
\\
& \times & L_{\mu\nu} w^{\mu\nu}(\bbox{p},\bbox{p}_N) S(p,{\cal E}) 
\frac{k'}{2} \frac{p_N}{2}\ .
\label {eq:d4sigma}
\end{eqnarray}
{}From Eq.~(\ref{eq:dsigma5}) or (\ref{eq:d4sigma}) one obtains the $t$-channel
($u$-channel) inclusive cross section by integrating over the undetected
nucleon (lepton) momentum.  
In so-doing, it is convenient to convert to integrals over the variables $p$ and
${\cal E}$.
To this end, we first make use of the fact that $\bbox{p}_N=\bbox{q}+\bbox{p}$ 
and ${\bbox{k}}^{\prime}=\bbox{u}+\bbox{p}$, so that for fixed $\bbox{q}$ one 
has $d^3 p_N=d^3p$ and for fixed $\bbox{u}$ one has $d^3k'=d^3p$. 
Second, we write $d^3p=d\phi d\cos\theta p^2dp$, where $(\theta,\phi)$ are
defined with respect to $\bbox{q}$ in the case of $t$-channel scattering and 
with respect to $\bbox{u}$ in the case of $u$-channel scattering. 
Third, we make use of the $({\cal E}, {\bbox{p}})$ relations in 
Eqs.~(\ref{eq:eps}) and (\ref{eq:emax}) to transform from $d\cos\theta$ to 
$d{\cal E}$, getting
\begin{equation}
  d\cos\theta =
    \left\{
    \begin{array}{cl}
      (E_N/pq)d{\cal E}\ , & \mbox{$t$-channel} \\
      (\epsilon'/pu)d{\cal E}\ , & \mbox{$u$-channel}
     \end{array}
     \right.
\end{equation}
respectively.
Finally, we obtain for the cross section
\begin{eqnarray}
  d\sigma & = & (2\pi)\int_{\cal D} d\phi\int pdp\int d{\cal E} S(p,{\cal E})
    \left(\frac{1}{16\epsilon E}\right) g^4 D_V(Q^2)^2 
\nonumber
\\
& \times & L_{\mu\nu} w(\bbox{p},\bbox{p}_N)^{\mu\nu}\frac{dQ_f}{\rho} \ ,
\label{eq:dsigma6}
\end{eqnarray}
where
\begin{mathletters}%
\begin{eqnarray}
  \frac{dQ_f}{\rho} &=& \frac{1}{q}\frac{d^3k'}{(2\pi)^3 2\epsilon'}\ ,
    \mbox{\ \ \ \ \ \ $t$-channel}\\
\label{eq:dqf1}
  \frac{dQ_f}{\rho} &=& \frac{1}{u}\frac{d^3 p_N}{(2\pi)^3 2 E_N}\ ,
    \mbox{\ \ \ \ \ $u$-channel} \ ,
\label{eq:dqf2}
\end{eqnarray}
\end{mathletters}%
${\cal D}$ denoting the allowed region of integration in the $({\cal E}, p)$
plane (for the case of electrons, we have assumed the extreme relativistic limit
($k=\epsilon$)).

{}From the above, the explicit expressions
\begin{eqnarray}
  \frac{d^2\sigma}{d\Omega' d\epsilon'}  & = &
    \frac{1}{2\pi} \frac{1}{32} \frac{k'}{\epsilon} \frac{1}{q} g^4 D_V(Q^2)^2 
\label {eq:d2sigmat}
\\
& \times &    \int_{\cal D} p dp \int \frac{d{\cal E}}{E} S(p,{\cal E})
    \overline{L_{\mu\nu} w^{\mu\nu}(\bbox{p},\bbox{p}_N) }
\nonumber
\end{eqnarray}
and
\begin{eqnarray}
&&  \frac{d^2\sigma}{d\Omega_N d E_N}  = 
    \frac{1}{2\pi} \frac{1}{32} \frac{p_N}{\epsilon} \frac{1}{u} g^4 
\nonumber
\\
&& \times 
    \int_{\cal D} p dp \int \frac{d{\cal E}}{E} S(p,{\cal E}) 
    \overline{L_{\mu\nu} w^{\mu\nu}(\bbox{p},\bbox{p}_N)} D_V(Q^2)^2 
\label {eq:d2sigmau}
\end{eqnarray}
for the $t$- and $u$-inclusive cross sections, respectively, follow.
Note that the azimuthal integration has already been carried out,
introducing an azimuthally averaged single-nucleon tensor (or, equivalently, 
single-nucleon cross section) as done, for example, in 
Refs.~\cite{Day,Cen96}.

The formula given in Eq.~(\ref{eq:d2sigmat}) constitutes the conventional 
starting point for the analysis of inclusive QE $(e,e')$ scattering in the 
PWIA.\footnote[3]{Typically, one sees integrations over $E$ rather than 
${\cal E}$ in the literature.} 
Previously reported treatments of QE $(\nu, N)$ scattering, however, have not
made use of this framework\cite{Gar93b,Hor93}. 
The advantage of writing $d\sigma(\nu, N)$ in the form of 
Eq.~(\ref{eq:dsigma6}) is two-fold: First, it makes explicit the dependence of 
the QE cross section on the experimentally-fixed kinematic variables 
$(u_0, {\bbox{u}})$ via the specification of the integration region ${\cal D}$ 
and the appearance of $u$ in the integration measure $1/\rho$. 
Second, it makes the role of the one-body nuclear spectral function explicit,
thereby making the nuclear model-dependence of the cross section more
transparent.  

When the single-nucleon tensor $w^{\mu\nu}(\bbox{p},\bbox{p}_N)$ refers to an 
on-shell nucleon, the expressions in Eqs.~(\ref{eq:dsigma6}) and 
(\ref{eq:d2sigmat}) 
carry a dependence on the free-nucleon form factors. For general neutral current
scattering, three such form factors contribute: $G_i(Q^2)$, where $i=E,M,$ or 
$A$ denotes the Sachs electric and magnetic form factors and the
axial-vector form factor, respectively\cite{Mus94}. 
In the case of $t$-channel scattering, one is then able to employ 
Eq.~(\ref{eq:dsigma6}) to extract information on the single-nucleon form factors
at a single value of $Q^2$. Since $t$ is fixed experimentally in this case, one 
may factor the form factors out of the integral over $({\cal E}, p, \phi)$, 
leaving only an integral over the spectral function and various kinematic 
factors resulting from the contraction of $L_{\mu\nu}$ and $w^{\mu\nu}$. To the
extent that one's choice of $S(p,{\cal E})$ is realistic and that the PWIA is 
valid, one obtains a more or less reliable determination of the $G_i(Q^2)$. 
An important feature of $t$-channel scattering is that $q$, $\omega$ and the 
electron scattering angle $\theta_e$ can all be fixed in an inclusive 
measurement, allowing (in the plane-wave Born approximation) the various 
longitudinal ($L$), transverse ($T$), $\ldots$ hadronic responses to be 
separated before attempting to determine the $G_i(Q^2)$, to be contrasted 
with the situation discussed below for $u$-channel inclusive scattering.

In contrast, even were the on-shell approximation to be a good one, 
$u$-channel neutrino scattering does not allow one to
extract the $G_i$ at a single value of $Q^2$. 
Since $U^2$ rather than $Q^2$ is fixed, the value of $Q^2$ varies as one 
integrates over the allowed region in $({\cal E}, p)$ space. 
For example, for kinematics typical of the LSND experiment (see 
Fig.~\ref{fig:Fig3}), $Q^2$
varies over the range (see next Section) $0\leq|Q^2|\leq 0.06$ 
$(\hbox{GeV/c})^2$ as ${\cal E}$ and $p$ vary over the allowed region 
${\cal D}$ (allowing the neutrino scattering angle to vary over all possible 
values). Similarly, for kinematics typical of the BNL experiment\cite{Ahr87}, 
$\epsilon = 1.3$ GeV and, say, $T_N = 500$ MeV, the corresponding range is 
$0\leq |Q^2|\leq 2.3$ $(\hbox{GeV/c})^2$. 
In either case, the form factors associated with the struck nucleon contribute
to the $u$-channel cross section over a range of $Q^2$. In order to extract
information about the form factors from this cross section, one is forced to
adopt some parameterization for the $Q^2$-dependence of the form factor
and fit the parameters to the measured cross section. Furthermore, as 
alluded to above, in $u$-channel inclusive scattering the values of $q$, 
$\omega$ and the neutrino scattering angle $\theta_\nu$ all vary when 
performing the integrations and thus no separation into isolated $L$, $T$, 
$\ldots$ hadronic responses is possible. 

All of these aspects of 
$u$-channel inclusive scattering stand in contrast with the $t$-channel
situation discussed above in which a parameterization-independent form factor
determination is possible. Only in the case of elastic scattering from $A=1$ 
targets are $u$- and $t$-channel processes equivalent. In the latter case, 
one has
\begin{equation}
  (K+P_A)^2+Q^2+U^2=2m^2_N+2{m_\ell^2}\ \ \ ,
\label{eq:stu}
\end{equation}
where ${m_\ell}$ is the mass of the lepton and 
$(K+P_A)^2=(\epsilon+m_N)^2-\bbox{k}^2\approx m^2_N-2\epsilon m_N$ in the lab 
frame. For scattering from a single-nucleon target, then, specifying $U^2$ is 
equivalent to specifying $Q^2$. Consequently, $u$-channel scattering can be 
used to perform a parametrization-independent form factor determination in this
case.

Finally, as discussed in Sec.~\ref{sec:nuclmodels} and the Appendix, another 
issue to be confronted is the fact that in general the struck nucleon is 
initially off-shell and accordingly the 
relationship to on-shell single-nucleon form factors is not obvious. In the 
present work we use a generalization of the popular $cc1$ prescription of 
de Forest \cite{Deforest} when modeling this type of behaviour. At least 
within the context of this approach we shall see that this model dependence 
is rather weak for the observables of interest (see Sec.~\ref{sec:axialff}).

\section{ RFG predictions for $\lowercase{\bbox{u}}$-inclusive scattering}
\label{sec:RFG}

Although already considered in Ref.~\cite{Hor93}, we wish to revisit in this
Section the RFG predictions for the $u$-inclusive scattering partly to
complement the findings of that work, which are rederived here via
the alternative route of integrating in proper regions of the $({\cal E},p)$ 
plane, and partly to establish analytic expressions for some general features of
the RFG $u$-inclusive cross section not presently available in the literature to
our knowledge. 

As discussed briefly in the previous Section, while in the $t$-channel it 
turns out to be possible to obtain a compact,
simple expression for the RFG inclusive cross sections or, equivalently, for the
longitudinal $R_L$ and the transverse $R_T$ response functions, this is not so
in the $u$-channel.
The reason is precisely the one mentioned at the end of the previous Section,
namely, the $Q^2$-dependence arising from the presence of the single-nucleon
form factors. For the $t$-inclusive case, $Q^2$ is fixed so that the
nucleon form factors may be factored out of the integrals over ${\cal E}$ and
$p$. By contrast in the $u$-channel $Q^2$ varies over 
the integration region ${\cal D}$; hence the nucleon's form factors cannot be 
brought out of the integrals except (and then only approximately) in special 
kinematic situations where the 
$G_i(Q^2)$ vary gently over ${\cal D}$.

To provide an appreciation for the behavior of $Q^2$ in the $({\cal E},p)$ 
plane we display its variation in Fig.~\ref{fig:Fig5} for the same kinematical 
situation of Fig.~\ref{fig:Fig3}, considering, as an example, the 
instance of $\bbox{p}_N$ and $\bbox{k}$ being parallel.
The explicit dependence upon ${\cal E}$ and $p$ of $Q^2$ turns out to be
\begin{eqnarray}
  Q^2 & = & \left(T_N+E_S+{\cal E}\right)^2
\label{eq:q2par}
\\
& + & \frac{k}{|k\mp p_N|}\left[\left(k-T_N-E_S-{\cal E}\right)^2
    +\left(k\mp p_N\right)^2-p^2\right]
\nonumber
\end{eqnarray}
where the upper (lower) sign corresponds to $\bbox{k}$ and $\bbox{p}_N$ being 
parallel (anti-parallel).
{}From Fig.~\ref{fig:Fig5}, it clearly appears that $Q^2$ varies rapidly with 
$p$ but only mildly with ${\cal E}$. 

Notwithstanding this feature we try to express analytically a few general
properties of the RFG $u$-inclusive cross section.
To this end, we recall~\cite{Cen96} that the RFG spectral function reads
\begin{eqnarray}
  S_{RFG}(p,{\cal E}) &=& 2 \theta(k_F-p)
\label{eq:SRFG}
\\
&\times&
\delta\left({\cal E}
    -\sqrt{k_F^2+m_N^2}+\sqrt{p^2+m_N^2}\right) \ ,
\nonumber
\end{eqnarray}
where the factor 2 accounts for the spin degeneracy.
Note that the above cannot be directly derived from the expression in
Eq.~(\ref{eq:S}), where the states are normalized according to (\ref{eq:pq}). 
Instead, one requires the normalization
\begin{equation}
  \langle\bbox{p}|\bbox{q}\rangle = 2 E_{\bbox{p}}\ \Omega\ 
    \delta_{\bbox{k},\bbox{p}}\ ,
\label{eq:pvbox}
\end{equation}
$\Omega$ being the (large) volume enclosing the Fermi gas.
Clearly, the support in Eq.~(\ref{eq:SRFG}) in the $({\cal E},p)$ plane is 
nonzero only along the curve
\begin{equation}
  {\cal E}^{RFG} = \sqrt{k_F^2+m_N^2}-\sqrt{p^2+m_N^2} \ .
\label{eq:Ssupp}
\end{equation}
It is then natural to expect the maximum of the $u$-inclusive cross section 
to correspond to the situation where ${\cal D}$ encompasses the whole of the 
RFG spectral function. In the strict RFG model this requires 
\begin{equation}
  I_u = \sqrt{k_F^2+m_N^2}-m_N = \epsilon_{F} - m_N = T_F > 0\ ,
\label{eq:Iuii}
\end{equation}
which insures that the piece of the boundaries of ${\cal D}$ stemming from 
${\cal E}^+$ does not cut off a section of the curve defined by
Eq. (\ref{eq:Ssupp}).

For kinematic conditions under which the $G_i(Q^2)$ vary mildly over the
integration region ${\cal D}$ and over variations in $(u_0, \bbox{u})$, it
follows from Eq.~(\ref{eq:Iuii}) that the maximum of the cross section will 
occur for the following kinetic energy of the outgoing nucleon:
\begin{equation}
  T_N^{max} = \frac{k \cos^2\theta_N}{1+m_N/(2k) + k/(2m_N)\sin^2\theta_N} \ .
\label{eq:TNmax}
\end{equation}
For purposes of comparison, we note that the same principle of maximum overlap
between the support of the spectral function and ${\cal D}$ leads to the 
following result for the maximum of the cross section: 

\begin{eqnarray}
  \omega_{max} & = & \sqrt{\bbox{q}^2+m_N^2} - m_N 
  \Rightarrow
\nonumber
\\
\omega_{max} & = & |Q^2_{max}|/2m_N\equiv [q^2-\omega_{max}^2]/2m_N\ ,
\label{eq:ommax}
\end{eqnarray}
in the $t$-channel (see Ref.~\cite{Cen96}). 
Actually Eq.~(\ref{eq:ommax}) holds exactly for properly reduced RFG response
functions where the single-nucleon form factor dependences are removed 
(see Ref.~\cite{Alb88}). If not so reduced, then at high momentum transfer the
prediction in Eq.~(\ref{eq:ommax}) is appreciably altered by the single-nucleon
physics. For the $u$-channel case, we find 
that the expression in Eq.~(\ref{eq:TNmax}), which provides the maximum for the
cross section, is valid
only for large $k$ and small $\theta_N$, where it turns out that the 
single-nucleon physics has less of an impact.

A novel feature of the RFG $u$-inclusive cross section, with respect to the
$t$-inclusive one, is its unexpected vanishing in some range of $T_N$. 
To illustrate this feature, we recall that the $u$-inclusive cross section is 
fixed by three parameters: $k$, $p_N$ and $\theta_N$. 
Now for $p^{-}_{<}=u-\delta > k_F$ no overlap exists between ${\cal D}$ and 
the curve of 
Eq. (\ref{eq:Ssupp}): Hence the vanishing of the cross section. 
This situation clearly corresponds to a negative value of the intercept $I_u$. 
However, when the intercept $I_u$ is positive, the cross section might also 
vanish, provided that $\delta-u$ is larger than $k_F$. 
One is thus led to consider the equation                    
\begin{equation}
  u-\delta=k_F \ ,
\label{eq:d-u}
\end{equation}
which admits two real, positive roots, namely
\begin{equation}
  T_N^{(1,2)} = 
    \frac{2 m_n k^2\cos^2\theta_N-\alpha(2 k-\alpha)(k+m_N-\alpha)
    \pm k|\cos\theta_N| \sqrt{\Delta}}
    {2(k+m_N-\alpha+k\cos\theta_N)(k+m_N-\alpha-k\cos\theta_N)}
\label{eq:TN12}
\end{equation}
with
\begin{eqnarray}
  \Delta &=& 4m_N^2 k^2 \cos^2\theta_N 
\nonumber
\\
&-& \alpha(2k-\alpha)(2m_N-\alpha)
    (2k+2m_N-\alpha)
\label{eq:Discr}
\end{eqnarray}
if, and only if,
\begin{equation}
  \cos\theta_N\geq\frac{\sqrt{\alpha(2m_N-\alpha)(2k-\alpha)
    (2k-\alpha+2m_N) }}{2m_Nk}\ ,
\label{eq:costh}
\end{equation}
where $\alpha=E_S+k_F$ and
\begin{equation}
  E_S = m_N - \sqrt{k_F^2+m_N^2} \approx -\frac{k_F^2}{2 m_N}
\label{eq:ES}
\end{equation}
is the {\em negative} separation energy of the RFG (see Ref.~\cite{Cen96}).

It thus appears that a critical angle, $\theta_0$, exists such that for
$\theta_N\leq\theta_0$ the $u$-inclusive cross section vanishes in the range
$T_N^{(1)}\leq T_N\leq T_N^{(2)}$. 
The asymptotic value of $\theta_0$ for large lepton momenta $k$ is $52^0$ at 
$k_F$ = 225 MeV/c and $\theta_0$ exists only for $k\geq k_F$.
Note that when $k=k_F$ then $\theta_N=0^0$ and the two roots in 
Eq.~(\ref{eq:TN12}) coincide, namely
\begin{equation}
  T_N^{(1)} = T_N^{(2)} = - E_S\ .
\label{eq:tn0}
\end{equation}
For $k>k_F$ the maximum distance between the roots in Eq.~(\ref{eq:TN12}) 
(the maximum
range over which the cross section vanishes) occurs for $\theta_N=0$. Then it
gradually decreases as $\theta_N$ is increased from $0^0$ to $\theta_0$.

A final consideration relates to the traditional handling of Pauli blocking 
in the RFG. In the $t$-channel, this blocking gives rise to the experimentally 
unsupported linear energy behaviour of the cross section at low $\omega$ 
when $q<2k_F$. Consequently, one does not expect the RFG cross section
to be credible until $q>2k_F$.
In the $u$-channel, Pauli correlations may also be incorporated, 
although one may debate the appropriate method for treating them. 
At the crudest level of approximation, for example, one may model
these correlations simply by requiring that the cross section vanish 
for $p_N<k_F$ ($T_N<T_F$). Importantly for comparisons with the LSND 
measurements, an experimental cut is made at $T_N=60$ MeV
($p_N\approx 340$ MeV), lessening the impact of such modeling.

Let us now conclude this Section by presenting some typical $u$-inclusive 
neutrino cross sections using the RFG.
For this purpose we first recall that the RFG is a symmetric system with equal
number of protons and neutrons (Z=N=A/2), all of them on the mass-shell.
Then we insert the spectral function of Eq.~(\ref{eq:SRFG}) into 
Eq.~(\ref{eq:d2sigmau}),
yielding
\begin{eqnarray}
  \frac{d^2\sigma}{d\Omega_N d E_N}  & = &
    {\cal N} \frac{1}{\rho} \frac{1}{(2\pi)^4} \frac{1}{16} 
    \frac{p_N}{u} \left(\frac{g}{M_V}\right)^4 
\label {eq:d2RFG}
\\
& \times &
    \int_{\cal D} dp \, p \int d{\cal E} \frac{S_{RFG}(p,{\cal E})}{E}
    L_{\mu\nu} w^{\mu\nu}(\bbox{p},\bbox{p}_N) 
\nonumber
\end{eqnarray}
where $\rho = 2 k_F^3/(3\pi^2)$ is the Fermi gas density and ${\cal N}$
indicates either the proton (Z) or the neutron (N) number.
Furthermore, when applied to neutrino scattering, the coupling 
to mass ratio which enters is
$g/(4 \cos\theta_W M_W)=\sqrt{G_F/(2\sqrt{2})}$, $G_F$ being
the Fermi constant (we have assumed the $\rho$ parameter to be
equal to unity for simplicity), and\cite{Hor93}
\begin{equation}
  L_{\mu\nu} w^{\mu\nu}(\bbox{p},\bbox{p}_N) =
   2 [ V_{VV} +V_{VA} +V_{AA} ] \ ,
\label{eq:lw}
\end{equation}
where explicit expressions both for the on- and off-shell 
quantities in Eq.~(\ref{eq:lw}) are given in terms
of the weak neutral current form factors in the Appendix.

Since our focus in this work is on $^{12}$C, we take $k_F=225$ MeV/c. 
The results of our calculations are displayed in Fig.~\ref{fig:Fig6}a, 
where we again consider kinematics typical of the LSND experiment 
($\epsilon$ = 200 MeV) and, for purposes of illustration, we explore two 
different orientations between the incoming neutrino and the outgoing nucleon 
(either proton or neutron), namely $\theta_N=20^\circ$ and $60^\circ$. 
We observe the following:
\begin{itemize}
\item[i)]{the cross section decreases with the angle $\theta_N$ for 
low values of $k$, at least for not too large values of 
$\theta_N$;}
\item[ii)]{the neutron cross section is larger than the proton
cross section.}
\end{itemize}

In Fig.~\ref{fig:Fig7}a the neutrino cross sections for the RFG 
(but with only outgoing protons)
are calculated using kinematics corresponding to the average energy of the
Brookhaven neutrinos ($\epsilon$ = 1.3 GeV). In this case, the experimental 
cut is made at $T_N=$ 200 MeV and so Pauli blocking plays a minor role. 
In contrast to the situation for low-energy neutrinos, 
the cross sections increase with $\theta_N$. In addition, the previously
noted possibility of a vanishing $u$-inclusive cross section for certain
values of $\theta_N$ appears in Fig.~\ref{fig:Fig7}a: 
A sizable gap in the cross section occurs for 
$\theta_N=20^0$, but not for $\theta_N=60^0$, since the value of the limiting 
angle in this case turns out to be $\theta_0=25.7^0$.

\section{ HM predictions for $\lowercase{\bbox{u}}$-inclusive scattering}
\label{sec:nuclmodels}

We next turn to the hybrid harmonic oscillator shell model\cite{Cen96} and 
its extensions to inclusion of a spreading width. 
While these choices clearly do not exhaust the list of possibilities --- in
particular in this work we do not deal with the final-state dynamics 
issues that were treated in Ref.~\cite{Gar93b} ---
we are able to set a scale for the theoretical uncertainty in the 
extracted value of $G_{\scriptscriptstyle{A}}^{(s)}$, as discussed in the next 
Section.

The HM developed in Ref.~\cite{Cen96} differs from the RFG in that it has 
the states below the Fermi surface confined, therefore requiring the 
struck nucleon to be off-shell (see the Appendix). In the PWIA 
the states above the Fermi surface are taken to be plane waves.
Accordingly in this model one accounts for the confinement of the nucleons in
the initial state, which is of course important for a realistic description of
low-energy neutrino-nucleus inelastic scattering, however at the expense of 
losing the Lorentz covariance characteristic of the RFG.
In other words the hybrid model is only semi-relativistic.

In principle, this confinement is generated by a Hartree-Fock well; 
in practice, as a simple, tractable model we use a harmonic oscillator 
as the confining potential, namely
\begin{equation}
  V(r) = \frac{1}{2} m_N \omega_0^2 r^2 - \bar{V} \ ,
\end{equation}
$\bar{V}$ being a positive quantity fixed in such a way to reproduce the
experimental separation energies, $E_s=15.96$ (18.72) MeV for protons 
(neutrons). For the oscillator frequency, we use 
$\omega_0=41/A^{1/3}$ MeV. The HM spectral function is then easily found to be 
\begin{eqnarray}
  S^{HM}(p,{\cal E}) &=& \sum_{N=0}^{N_{MAX}}
    \delta[{\cal E} - (N_{MAX}-N)\omega_0 + \Delta E]n_N(p) \nonumber\\
  &=& \sum_{N=0}^{N_{MAX}} \delta[E-(N+\frac{3}{2})\omega_0
    - \bar{V} - m_N]n_N(p) 
\nonumber
\\
\label{eq:S-HM}
\end{eqnarray}
where the shift in the excitation energy ${\cal E}$ stemming from the binding of
the nucleons is denoted by $\Delta E$.
In the above, the momentum distribution for a given shell identified by
the quantum number $N=2n+\ell$ is given by
\begin{equation}
  n_N(p) = 2 \sum_{n\ell}\frac{2\ell+1}{4\pi}|\varphi_{n\ell}(p)|^2 \ ,
\end{equation}
where $\varphi_{n\ell}(p)$ are the harmonic oscillator radial wave functions 
in momentum space. For simplicity (and in good accord, for example,  with 
mean-field descriptions of light $N=Z$ nuclei) we assume the momentum 
distributions of protons and neutrons in a given shell to be the same. 
In contrast with the RFG spectral function discussed above one finds that
the HM spectral function in Eq.~(\ref{eq:S-HM}) is nonzero on a set of 
parallel lines in the (${\cal E},p$) plane, one for each 
shell ({\em e.~g.,} two lines for $^{12}$C), at variance with the case of the 
RFG, whose spectral function is nonzero on the single curve given by 
Eq.~(\ref{eq:Ssupp}). As mentioned in the Introduction, in the present work 
when considering the HM we employ only a harmonic oscillator basis with its 
overly strong confinement to provide the strongest contrast with the RFG model 
--- a sort of ``worst-case'' scenario.

As one straightforward extension of the above formalism where specific 
classes of correlations effects may be modeled we also allow the 
single-particle states to acquire a spreading width. 
This extension is implemented by adding a complex self-energy 
to the single-particle energies, thus defining the fermion propagator as
\begin{equation}
  G_{HM}^{(SW)}(p,E) = \sum_{N=0}^{N_{MAX}} 
    \frac{n_N(p)}{E-[\epsilon_N+\Sigma(\epsilon_N)-\epsilon_F]} \ ,
\end{equation}
where $\epsilon_N=m_N+(N+\frac{3}{2})\omega_0-\bar{V}$ represents the
energy of the nucleon in the $N$th shell.
The self-energy $\Sigma$, of course, arises from second- (or higher-) order 
insertions on the fermion propagation lines: Here, we adopt the 
point of view of parametrizing it in terms of the following function
\begin{equation}
\Sigma(\epsilon) = \Delta(\epsilon) - i \frac{\gamma(\epsilon)}{2} \ ,
\end{equation}
with
\begin{equation}
  \gamma(\epsilon) = 2\alpha\frac{\epsilon^2}{\epsilon^2+\epsilon^2_0}
    \frac{\epsilon^2_1}{\epsilon^2+\epsilon^2_1}\theta(-\epsilon) \ ,
\label{eq:gamma_sw}
\end{equation}
whereas $\Delta(\epsilon)$ is obtained from $\gamma(\epsilon)$ via a subtracted
dispersion relation. We follow Ref.~\cite{Mah81,Smi88} in setting 
$\alpha=10.75$ MeV, $\epsilon_0=18$ MeV and $\epsilon_1=110$ MeV.
The spectral function is proportional to the imaginary part of the hole
propagator, $S_{HM}^{(SW)} = \text{Im} G_{HM}^{(SW)}/\pi$: Hence, in principle, 
the spreading width extends the support of the spectral function to the whole 
first quadrant of the (${\cal E},p$) plane; in practice, $S_{HM}^{(SW)}$ 
remains concentrated around the lines where the spectral function for a pure 
harmonic oscillator is non-vanishing.

We turn now to an examination of some of the model dependences that we obtain. 
In Figs.~\ref{fig:Fig6} and \ref{fig:Fig7}, we display 
the neutrino cross sections for the HM and its extended version with a 
spreading width for the same kinematical conditions explored above using 
the RFG. One sees from the figures that, although the basic features of the 
RFG cross sections studied in Section \ref{sec:RFG} do not appear to be much 
altered by the nucleons' confinement, yet some differences do show up. In 
particular, the peaks of the cross sections in the HM appear shifted and 
the phenomenon of the vanishing of the cross section, which is characteristic 
of the RFG, is still present in the hybrid model, although in a narrower range 
of energies. Note also in comparing panels b) and c) that the effect of the 
spreading width, and therefore this particular class of correlation 
effects, is negligible and is henceforth ignored. 

While the nuclear modeling used in these comparisons of RFG and HM results 
can obviously be extended in many directions, including the use of alternative 
mean-field bound wave functions and incorporation of final-state 
interaction effects, the theoretical uncertainties are likely reasonably 
characterized by the present ``extreme'' situations. Each model has its 
own merits. In particular, the RFG maintains covariance, which becomes 
more relevant for the BNL kinematics, whereas the HM brings in the 
confinement of the (off-shell) struck nucleon, albeit at the expense of 
covariance. Since the former involves only plane-wave nucleons while the 
latter has rather strongly confined HO wave functions for the struck 
nucleons, most other choices might be expected to fall somewhere between 
the results presented here. If the span of predictions given in the 
figures for the double-differential cross section is taken at face value, 
then substantial nuclear model dependences are anticipated and any hope 
of using absolute cross sections for the purpose of learning about 
axial-vector strangeness in the nucleon is correspondingly low. 
As we discuss in the next section, however, there does appear to be a 
strategy whereby one can mitigate this model-dependence in the extraction of
$\Delta s$ from neutrino observables.

\section{ Strange axial-vector form factor}
\label{sec:axialff}

\def\gass{{g_A^{(s)}}}

We now address the use of neutrino scattering to probing nucleon strangeness.
To this end, we recall that the nucleon's NC axial-vector form factor may 
be written as \cite{Mus94,Mus92} (see the Appendix)
\begin{equation}
  \widetilde{G}_{A} = \left[\xi_A^{T=1} G_A^{(3)}\tau_3+ 
	\xi_A^{T=0} G_A^{(8)} + \xi_A^{(0)} G_A^{(s)}\right] \ ,
\label{eq:axialff}
\end{equation}
where
\begin{mathletters}%
\begin{eqnarray}
  G_A^{(3)} &=& (1/2)(D+F)G_D^A(\tau) \\
  G_A^{(8)} &=& (1/2\sqrt{3})(3F-D) G_D^A(\tau)\\
  G_A^{(s)} &=& g_A^{(s)} G_D^A(\tau)\\
  D+F	    &=& 1.257
\end{eqnarray}
\label{eq:axialffb}
\end{mathletters}%
\noindent are SU(3) octet axial-vector form factors, $D$ and $F$ are the 
associated
SU(3) reduced matrix elements, $g_A^{(s)}\equiv\Delta s$ is the strange
quark contribution to the nucleon's axial charge, and $\tau=|Q|^2/4m_N^2$. 
The coefficients $\xi_A^{(a)}$
are determined by the axial-vector coupling of the $Z^0$ to the quarks,
and take on, at tree level in the Standard Model, the values $-2$, $0$, and $1$ 
for $a$ being $T=1$, $T=0$, and $0$, respectively. For neutrino scattering,
electroweak radiative corrections are of ${\cal O}(\alpha/4\pi)$ 
\cite{Mar80,Mus90},
so for purposes of this analysis, we may employ the tree-level values. Under
this approximation, any isoscalar component of the nucleon's axial-vector 
form factor arises solely from the strange quark term. 

The quantity
$G_D^A(\tau)$ is a dipole parameterization for the $Q^2$-dependence of
the axial-vector form factors. In principle, one has no rigorous justification
for choosing a dipole form and for assuming that the different form factors in
Eq.~(\ref{eq:axialffb}) display the same $Q^2$-behavior. In the case of
high-energy neutrino scattering, where values for $|Q^2|$ on the order
of one $(\hbox{GeV}/c)^2$ are achieved, these assumptions can have rather
drastic consequences for one's determination of $\Delta s$. As indicated
in Refs.~\cite{Ahr87,Gar93a}, the value of $\Delta s$ is strongly correlated
with the dipole mass parameter, $M_A$. As far as a high-energy neutrino
scattering determination of $\Delta s$ is concerned, 
the impact of choosing a different
parameterization or of allowing the octet and singlet form factors to
display different $Q^2$-behavior is unknown. At kinematics relevant to
the LSND experiment, the issue of non-leading $Q^2$-dependence is much
less serious. In this case, one may view the dipole form factors as a way
to include the first derivative-term in a Taylor expansion of the form factor
(with respect to $Q^2$). Moreover, as shown in Refs.~\cite{Mus94,Hor93},
the correlation between $\Delta s$ and $M_A$ is negligible. Thus, we may
safely employ the dipole parameterization for purposes of analyzing 
low-energy scattering without introducing significant uncertainty.

	To illustrate the relationship between $\Delta s$ and the choice of
nuclear model, we first plot in Fig.~\ref{fig:Fig8}
the differential cross section as a function
of outgoing nucleon kinetic energy for two different values of $\Delta s$:
0 and $-0.2$. The latter value corresponds roughly to the largest value
(in magnitude) of $\Delta s$ derived from the deep inelastic measurements.
The results for kinematics typical of the LSND experiment are shown in
Figs.~\ref{fig:Fig8}a,b while those typical of the BNL measurements are given 
in Fig.~\ref{fig:Fig8}c.
We note that in the case of the LSND experiment, for which the target is
a CH$_2$ based scintillator and for which the incident neutrinos are produced
by the decay of pions in flight, a cut on $T_N$ of $> 60$ MeV guarantees 
that only nucleons knocked out of carbon nuclei are detected \cite{Gar92}.
For these energies, one sees from Figs.~\ref{fig:Fig8}a,b that the nuclear 
model-dependence
is sufficiently large to prevent an unambiguous extraction of $\Delta s$.
For example, in the case of the proton knockout cross section, the difference
between the RFG prediction for $\Delta s = 0$ and the HM prediction for
$\Delta s = -0.2$ is comparable to or smaller than the difference between the 
RFG predictions
with $\Delta s = 0$ and $\Delta s = -0.2$. This feature does not appear to
persist at higher-energies (Fig.~\ref{fig:Fig8}c) where the model-dependence
for a given value of $\Delta s$ is significantly smaller than the dependence
on the value of $\Delta s$.

The authors of Ref.~\cite{Gar92} proposed that instead of analyzing the
proton and neutron cross sections separately, one ought to consider the
{\it ratio} of total proton to neutron yields. In contrast to the 
situation with the individual cross sections, the yield ratio is insensitive
to one of the primary experimental uncertainties -- the normalization of the
incident neutrino flux. Moreover, it was shown in Ref.~\cite{Gar93b} that the
yield ratio is less sensitive to final-state interactions than are the
individual cross sections. Thus, the use of the $p/n$ ratio would appear
to minimize the impact of experimental and theoretical uncertainties on
the extraction of $\Delta s$ from QE neutrino scattering data. In what
follows, we illustrate the degree to which the use of this ratio can also
reduce one's sensitivity to the choice of nuclear model. We
work with the ratio of cross sections rather than yields, assuming as in
Ref.~\cite{Gar92}
that the incident fluxes are peaked at some energy $\bar\epsilon$
and that working with the cross sections at $\epsilon=\bar\epsilon$ is 
sufficient. To that end, we first show in Fig.~\ref{fig:Fig9} 
the ratio of differential proton knockout and
neutron knockout cross sections as a function of $T_N$. {}From these curves,
we deduce that the spread due to the choice of nuclear model for a given
value of $\Delta s$ and $T_N$ is roughly 20\% or less than the change in
the ratio obtained by varying $\Delta s$ from zero to $-0.2$.

	In Fig.~\ref{fig:Fig10} we give the total cross sections, integrated 
over outgoing nucleon energy (with $T_N > 60$ MeV in the case of low-energy
neutrinos and $T_N > 200$ MeV for high-energy neutrinos)
as functions of $\Delta s$. In the case of the 
LSND experiment, it is the total proton and neutron yields that will be
obtained, making these curves more directly relevant to the interpretation
of the experiment. From Fig.~\ref{fig:Fig10}a,b, one 
sees that the model-sensitivity of one's
extracted value of $\Delta s$ is non-trivial in the case of low-energy
neutrino scattering. For a given value of the total
cross section, the two would yield values of $\Delta s$ differing by about
$0.5$. Thus, we would take a reasonable nuclear theory error bar on the
value of $\Delta s$ extracted from the single cross section to be 
$\delta_{\hbox{nuc}}(\Delta s) = \pm 0.25$. This uncertainty is as large in 
magnitude as the value of $\Delta s$ determined from deep inelastic scattering. 
For higher energy scattering, however, the corresponding nuclear model
uncertainty is much smaller: $\delta_{\hbox{nuc}}(\Delta s)=\pm 0.015$ (see 
Fig.~\ref{fig:Fig10}c).

	Finally, in Fig.~\ref{fig:Fig11} we give the ratio $R_\nu$ 
of proton to neutron total 
cross sections for low-energy scattering. In this case, one finds that the
ratio is significantly less sensitive to the choice of nuclear model than
in the case of the actual cross sections themselves. Indeed, we would deduce
a nuclear theory uncertainty in $\Delta s$ of $\approx \pm 0.015$ 
from the curves in Fig.~\ref{fig:Fig11}. This uncertainty represents an order
of magnitude improvement from the uncertainty associated with the use of the
individual cross sections. 

One other issue involving model dependence was also briefly examined, namely, 
that of the on- versus off-shell current descriptions employed. We 
evaluated the HM results using the on-shell single-nucleon current and 
found that $R_\nu$ in that case is lower than the RFG curve by about the 
amount it is higher for the off-shell results shown in Fig.~\ref{fig:Fig11}. 
Of course, such a calculation is not strictly correct and should be used only 
as a rough indication of the sensitivity to the prescription for the 
current, since some off-shellness must be used for the HM as demanded by 
the kinematics of the reaction. In fact, the rather weak dependence found is 
reassuring and suggests that, at least for the specific kinematical 
choices that are relevant here, the issue of what particular single-nucleon 
current to employ is likely even less important than the nuclear model 
dependence. To be as conservative as possible, we will multiply 
by two the uncertainty deduced from Fig.~\ref{fig:Fig11} and take 
$\delta_{\hbox{nuc}}(\Delta s) = \pm 0.03$.

The reason for this reduction in nuclear model
sensitivity when using $R_\nu$ is straightforward to understand. 
At the kinematics typical
of the LSND experiment, the range of $Q^2$ accessed as one integrates over
the allowed region ${\cal D}(\theta_N)$ is small. Consequently, the nucleon
form factors and kinematic factors vary gently and, to a first approximation,
may be factored out of the integral in the expression of 
Eq.~(\ref{eq:d2sigmau}). One is
then left with a product of the squared single-nucleon invariant amplitude
and an integral of the one-body spectral function. Differences between model
predictions would then arise from different distributions of strength 
in the spectral function over the allowed domain in the $({\cal E}, p)$
plane. In the ratio of proton and neutron cross sections, the integrals
of the spectral functions should cancel assuming the proton and neutron
$S(p,{\cal E})$ are identical. In this case, the ratio is independent of nuclear
model and is determined only by the single-nucleon form factors. Any model
dependence which appears in the ratio would then arise from one of the
following sources: (i) the variation in form factors and kinematic factors
over the integration region ${\cal D}(\theta_N)$, so that the factorization
into a product of single-nucleon and many-body physics is not exact; (ii)
differences between proton and neutron spectral functions; (iii) contributions
going beyond the PWIA. In the case of low-energy neutrino
scattering from carbon for $T_N > 60$ MeV, one expects the effects (i) and
(ii) to be small. Indeed, our results indicate that effects of type (i)
generate roughly a 10\% variation in the ratio for a given value of 
$\Delta s$, which translates into the nuclear theory uncertainty quoted above.

In arriving at our estimate of the nuclear theory uncertainty, we make 
no pretense of having performed definitive, state-of-the-art nuclear model
calculations of the kind reported in Refs.~\cite{Gar92,Gar93b}. 
Rather, our goal
was to give a reasonable upper bound on the expected spread in model 
calculations. We anticipate that any series of more sophisticated
model calculations would yield extractions of $\Delta s$ from $R_\nu$
differing by no more than our value for
$\delta_{\hbox{nuc}}(\Delta s)$. From this standpoint, it is
instructive to compare
our RFG and HM results for this ratio with the mean field calculation 
reported in Ref.~\cite{Gar92}. For a given value of $\Delta s$, the
mean field results and RFG results for $R_\nu$
are almost identical, while the HM results differ by less then 10\%.
This agreement holds in spite of much larger differences which appear
when one compares the mean field, RFG, and HM predictions for the individual
differential cross sections and for the ratio of differential cross sections
as a function of $T_N$. We further believe that this robust nature of
$R_\nu$ will persist when one compares different model calculations which
not only reproduce experimental results for the inclusive $t$-channel QE
responses but also include FSI's in the $u$-channel case. It was demonstrated
in Ref.~\cite{Gar93b} that the inclusion of FSI's in the mean field
approach can increase the predicted value of $R_\nu$ by more than 10\%
over the mean field and RFG predictions. This increase would imply
a shift in the extracted value of $\Delta s$ by more
than 0.03. Thus, it appears that any realistic model which is used in
this extraction must include FSI's. Nevertheless, we would argue that
the {\em spread} in extracted values of $\Delta s$ 
corresponding to the use of different nuclear models which include FSI's
would remain smaller than $\delta_{\hbox{nuc}}(\Delta s)$. Indeed, an
important check on our first estimate of $\delta_{\hbox{nuc}}(\Delta s)$
would be to compare model calculations which incorporate FSI's as well as
to explore the impacts of correlations and charge symmetry breaking.

	Finally, we compare the nuclear theory uncertainty in the neutrino 
scattering determination of $\Delta s$ with the theory uncertainties associated
with the extraction of $\Delta s$ from measurements of the $g_1(x, Q^2)$
deep inelastic structure function. We first remind the reader that one does
not necessarily expect the values of $\Delta s$ obtained from these two
different processes to agree. It has long been known, but perhaps not
widely appreciated in the nuclear physics community, that $\Delta s$ is a
renormalization-scale dependent quantity. In deep inelastic scattering, 
this scale is typically taken to be the same as the mean $\sqrt{|Q^2|}$ of
the reaction (between $\sqrt{2}$ and $\sqrt{11}$ $\hbox{GeV}/c$), 
whereas the scale appropriate to the quasielastic neutrino value is 
somewhere below the charm
quark mass. The latter corresponds to the scale below which one includes only
the three lightest quarks in an effective axial-vector NC \cite{Kap88}.
It is often assumed that the QCD evolution of $\Delta s$ between the two
scales is rather gentle, based on leading-order perturbative calculations,
yet it is conceivable that non-perturbative effects could invalidate this
assumption \cite{Jaf90}. At present, one can make no definitive statements
regarding the evolution of $\Delta s$ between these two scales. 

	A separate issue pertaining to the use of polarized structure functions
to determine $\Delta s$ is the use of SU(3) symmetry. In the standard 
operator product analysis of the $g_1$-sum, one employs an SU(3)
parameterization of the octet axial-vector matrix elements $\langle N |
A_\mu^{(a)}(0)| N\rangle$ ($a=3,8$ refers to the respective components
of the octet) in order to derive a value for $\Delta s$ or $\Delta\Sigma$.
The latter denotes the nucleon's singlet axial charge or, in the quark-parton
framework, the total light quark ($u$, $d$, $s$) contribution to the 
nucleon spin. The value of $\Delta s$ determined
in this fashion is quite sensitive to the quantity
\begin{equation}
{2\over\sqrt{3}}\langle N|A_\mu^{(8)} (0)| N\rangle = {1\over 3}(3F-D)\ \ \ .
\label{eq:axialme}
\end{equation}
In an SU(3)-symmetric fit to hyperon semileptonic decays, the combination
of reduced matrix elements $3F-D$ takes on a value of about 0.6\cite{Dai95}.
Because it involves a cancellation between two quantities, this number
is rather sensitive to uncertainties in the fit as well
as to corrections arising from SU(3)-breaking in the octet matrix elements.
Several analyses have been carried out recently \cite{Jen91,Dai95,Ehr94,Lic95} 
in which the hyperon decays were re-fit with allowance for
SU(3)-breaking. While different schemes for the incorporation of SU(3)-breaking
were employed in each case, a general trend does emerge: SU(3)-breaking
may reduce the matrix element in Eq.~(\ref{eq:axialme}) by 50\% or 
more compared to its
SU(3)-symmetric value, with an uncertainty of comparable magnitude. Such a
reduction would imply a value of $\Delta s\approx 0$ from deep inelastic
data, in contrast to the current average of the measurements $\Delta s
\approx -0.1$. The authors of these studies caution that precise numerical
value for the SU(3)-breaking correction to the quantity in 
Eq.~(\ref{eq:axialme}) is not
highly reliable, due to the nature of the aforementioned cancellation.
To be conservative, then, we take the SU(3)-breaking uncertainty in this
matrix element to be 50\% of its SU(3)-symmetric value. The corresponding
uncertainty in the strange-quark axial charge is $\delta_{\hbox{DIS}}(\Delta s)
\approx\pm 0.1$, a value having the same magnitude as the present average
for $\Delta s$ under the assumption of good SU(3) symmetry in the hyperon
semi-leptonic decays. The scale of this uncertainty is consistent with the
scale of the error obtained with the heavy-baryon chiral perturbation theory
analysis of Ref.~\cite{Jen91} and with the range of estimates for
SU(3)-breaking corrections given in Refs.~\cite{Dai95,Ehr94,Lic95}. 
The presence of this SU(3)-uncertainty weakens
the standard conclusion drawn from
inclusive, polarized deep inelastic measurements that the strange quarks 
are polarized oppositely to the direction of the nucleon's spin and that
the magnitude of the $s\bar s$  contribution is about one-third the total
quark contribution, $\Delta\Sigma$.

	By way of comparison, we note that $\delta_{\hbox{nuc}}(\Delta s)$ is
about one-third as large as $\delta_{\hbox{DIS}}(\Delta s)$. Moreover, the
analysis of the QE neutrino cross sections does not suffer from the kind
of SU(3)-breaking corrections and uncertainties which enter the interpretation
of the $g_1$-sum results. Although the same problematic combination of
SU(3) reduced matrix elements $3F-D$ enters the decomposition of the
nucleon's axial-vector NC form factor 
(see Eqs.~\ref{eq:axialff}-\ref{eq:axialffb}),
its contribution is suppressed by the coupling $\xi_A^{T=0}$. The latter is
identically zero at tree level in the Standard Model and is on the order of
0.01 when one-loop electroweak corrections are included. From this standpoint,
then, it appears that the ratio $R_\nu$ 
provides a theoretically cleaner window on
the strange-quark contribution to the nucleon's spin than
does inclusive, polarized deep inelastic scattering. The primary limitation
in the former case appears to be experimental error.

\section{Conclusions}
\label{sec:concl}

The study of low- and intermediate-energy semi-leptonic processes has played
an important role in helping uncover the nature of fundamental interactions.
In the case of QCD, semi-leptonic scattering is poised to illuminate the
way the strong interaction is realized in the structure of the nucleon. As
with the use of semi-leptonic scattering to test the Standard Model and its
possible extensions, the efficacy of this probe of the nucleon's structure
is limited by the reliability with which one can compute theoretically, or 
determine experimentally, the other hadron and nuclear structure contributions
to the relevant observables. It is important, then, that where hadron structure
or many-body theory is brought to bear on the interpretation of such
observables, an attempt be made to quantify the theoretical uncertainty.

In the present work, we have attempted to provide a theoretical error bar
for the determination of $\Delta s$ from quasielastic neutrino NC scattering.
The nucleon's strange quark axial charge is of interest since it has the
quark-parton interpretation as giving the strange-quark contribution to the
spin of a polarized nucleon. A large value for this quantity would imply that
the non-valence quarks play a more central role in the low-energy
characteristics of the nucleon than implied by the highly-successful and
intuitively satisfying quark model. To the extent that one may extract
the nucleon's NC axial-vector form factor from quasielastic neutrino
scattering at low momentum transfer, one has a direct probe of $\Delta s$.
In the ideal situation, one would have in hand sufficient information from
quasielastic $(e,e'N)$ measurements to perform this extraction without
heavy reliance on theoretical calculations of the QE response. At present,
sufficient $(e,e'N)$ data are lacking, rendering the use of nuclear models
unavoidable. 

We have tried to argue that were one to rely on the individual $(\nu, N)\nu'$
cross sections measured by LSND, the nuclear theory error bar on $\Delta
s$ would be sufficiently large to render the results inconclusive. Fortunately,
the ratio of total proton to neutron yields ($R_\nu$)
appears to be much less sensitive
to the choice of nuclear model, reducing by nearly one order of magnitude
the nuclear theory uncertainty in $\Delta s$. Based on the study of some
simple nuclear models with quite distinct assumptions regarding the many-body
dynamics, we estimate the nuclear theory error to be $\delta_{\hbox{nuc}}(
\Delta s) = \pm 0.03$. We anticipate that the spread in values for $\Delta s$
obtained from $R_\nu$ and more sophisticated nuclear models will be smaller
than this value for the uncertainty. We further note that $\delta_{\hbox{nuc}}
(\Delta s)$ is about a factor of three smaller in magnitude than 
$\delta_{\hbox{DIS}}(\Delta s)$, where the latter is derived from recent 
analyses of SU(3)-breaking in the octet of baryon axial-vector matrix elements.
In contrast, the interpretation of QE neutrino cross sections is relatively
free from such SU(3)-uncertainties.

\vspace{0.25in}
{\bf Acknowledgements:} The authors from Torino would like to thank the INT in 
Seattle for the hospitality extended to them during one phase of the work 
on this project. We also gratefully acknowledge useful discussions with
A. Adelberger, M. Frank, W. Haxton, and C. Johnson.

\appendix
\section*{}
\label{sec:appendix}

The quantities appearing in Eq.~(\ref{eq:lw}) are given by the following
expressions:
\begin{eqnarray}
V_{VV} &=&2[V_{11}+V_{12}+V_{22}]  \label{eq51a} \\
V_{VA} &=&2[V_{A1}+V_{A2}]  \label{eq51b} \\
V_{AA} &=&2A  \label{eq51c}
\end{eqnarray}
using the nomenclature of Ref.~\cite{Hor93}. On-shell one has from that 
reference
\begin{equation}
    V_{11} =
 4\tilde F_1^2 (P\cdot K \ P_N\cdot K' + P_N\cdot K \ P\cdot K' - m_N^2 K\cdot
 K') 
\end{equation}

\begin{equation}
    V_{12} =
 -4\tilde F_1\tilde F_2 K\cdot K' \ (P_N-P)\cdot(K-K')
\end{equation}

\begin{eqnarray}
    V_{22} & = &
 \frac{2\tilde F_2^2}{m_N^2} K\cdot K' \ 
(P\cdot K \ P_N \cdot K + P\cdot K' \ P_N\cdot K' 
\nonumber
\\
& + & m_N^2 K\cdot K')
\end{eqnarray}

\begin{equation}
    A = 4 \tilde G_A^2 (P\cdot K \ P_N\cdot K' + P_N\cdot K \ P\cdot K' 
    + m_N^2 K\cdot K') 
\end{equation}

\begin{equation}
    V_{A1} = 8 \tilde G_A\tilde F_1 (P\cdot K \ P_N\cdot K' - P_N\cdot K \ P
    \cdot K')
\end{equation}

\begin{equation}
    V_{A2} = 4 \tilde G_A\tilde F_2 K\cdot K' \ (P+P_N)\cdot (K+K')\ ,
\end{equation}
where $P$ and $P_N$ are the initial and final four-momenta of the nucleon, $K$
and $K'$ the initial and final four-momenta of the neutrino, $\tilde G_A$ is the
axial-vector form factor of the nucleon whereas $\tilde F_1$ and $\tilde F_2$
are the Dirac and Pauli weak NC form factors of the nucleon, related to the
Sachs NC form factors through the expressions
$\tilde G_E = \tilde F_1-\tau \tilde F_2$
and
$\tilde G_M = \tilde F_1+\tilde F_2$, with $\tau=|Q^2|/4m_N^2$.
For the nucleon's form factors we have used the Galster parameterization 
(see Refs.~\cite{Mus92,Don92}). These can be combined as in 
Eqs.~(\ref{eq51a}--\ref{eq51c}) to obtain the following for the on-shell 
versions of the quantities in Eq.~(\ref{eq:lw}):
\begin{eqnarray}
V_{VV}^{on}&=&m_N^4 \left\{ \left( {{Y\cdot Z}\over{m_N^2}} \right)^2
  {\tilde W}_2 -16\tau ({\tilde G}_E^2 -\tau {\tilde G}_M^2)
  \right\} \label{XA1}\\
V_{VA}^{on}&=&m_N^4 \left\{ 16\left( {{Y\cdot Z}\over{m_N^2}} \right) \tau 
  {\tilde G}_M {\tilde G}_A \right\} \label{XA2}\\
V_{AA}^{on}&=&m_N^4 \left\{ \left[ \left( {{Y\cdot Z}\over{m_N^2}} \right)^2
  +16\tau (1+\tau) \right] {\tilde G}_A^2 \right\}\ . \label{XA3}
\end{eqnarray}
Here we have used the following four-vectors: $Q=K-K'=P_N-P$, $Y\equiv K+K'$ 
and $Z\equiv P_N+P$, where then $Q\cdot Y=0$ and for on-shell nucleons 
$Q\cdot Z=0$. Furthermore, we define as usual ${\tilde W}_2\equiv 
[{\tilde G}_E^2+\tau {\tilde G}_M^2]/(1+\tau)$.

Off-shell we use a generalization of the $cc1$ prescription introduced by 
de Forest \cite{Deforest} for electron scattering; in our case we now have 
analogous quantities involving both vector and axial-vector neutral 
currents. Defining, as usual, $P^\mu \equiv P_N^\mu -Q^\mu =(E,\mathbf{%
p)}$, letting $\bar P^\mu \equiv P_N^\mu -\bar Q^\mu =(\bar E,\mathbf{p),}$ 
with $\bar E\equiv \sqrt{m_N^2+p^2}$ as in Ref. \cite{Deforest}, and in 
addition defining $\bar Z^\mu \equiv P_N^\mu +\bar P^\mu ,$ so that $Q\cdot
Y=\bar Q\cdot \bar Z=0,$ we obtain 
\begin{eqnarray}
V_{VV}^{off} &=&\left[ \left( Q\cdot \bar Q\right) ^2-\left( \bar Q\cdot Y
\right)^2\right] \tilde G_M^2+4m_N^2 Q^2\tilde G_E^{\prime 2}  \nonumber \\
&+&\left[ \left( Y\cdot \bar Z\right) ^2-\left(
Q\cdot \bar Z\right) ^2\right] \tilde W_2^{\prime }  \label{eq51d} \\
V_{VA}^{off} &=&4\left[ \left( Q\cdot \bar Z\right) \left( \bar Q\cdot Y\right)
\right.
\nonumber
\\
&-&\left.\left( Q\cdot \bar Q\right) \left( Y\cdot \bar Z\right) \right] \tilde
G_M\tilde G_A  \label{eq51e} \\
V_{AA}^{off} &=&\left[ \left( Q\cdot \bar Q\right) ^2+\left( Y\cdot \bar Z
\right)^2\right.   \nonumber  \\
&-&\left. \left( Q\cdot \bar Z\right) ^2-\left(
\bar Q\cdot Y\right) ^2-4m_N^2Q^2\right] \tilde G_A^2.  \label{eq51f}
\end{eqnarray}
In Eq. (\ref{eq51d}) we have introduced the following additional 
single-nucleon form factors:
\begin{eqnarray}
\tilde G_E^{\prime } &\equiv &\tilde F_1-\bar \tau \tilde F_2  \label{eq51h}
\\
\tilde W_2^{\prime } &\equiv &\frac 1{1+\bar \tau }\left[ \tilde G_E^{\prime
2}+\bar \tau \tilde G_M^2\right]   \label{eq51i}
\end{eqnarray}
in close analogy with the above on-shell quantities (see also Ref.~\cite{Cab}).
The results given here for the half-off-shell currents revert to the
on-shell results in Eqs.~(\ref{XA1}--\ref{XA3}) when $P$ and ${\bar P}$ 
become the same, {\it i.e.,} when the kinematics force the nucleon on-shell.

\begin{figure}
\caption{ Feynman diagram for exclusive lepton scattering in the one boson 
exchange approximation.
  }
\label{fig:Fig1}
\end{figure}

\begin{figure}
\caption{ The domain ${\cal D}$ over which one integrates for the 
$t$-inclusive cross
section at typical values of momentum and energy transfer. Note the more complex
structure of the boundary when the intercept $I_t$ given by 
Eq.~\ref{eq:eps0} is positive.
  }
\label{fig:Fig2}
\end{figure}

\begin{figure}
\caption{ The same as Fig.~2 for the $u$-inclusive cross section: a) LSND 
kinematics ($\epsilon=$ 200 MeV, $T_N=$ 60 MeV); b) BNL kinematics 
($\epsilon=$ 1.3 GeV, $T_N=$ 60 MeV). The boundaries shown all involve 
${\cal E}^-$ except for line BF which involves ${\cal E}^+$. Note that when
the neutrino and outgoing nucleon momenta are antiparallel quite remote regions
of the $({\cal E},p)$ plane are explored.
  }
\label{fig:Fig3}
\end{figure}

\begin{figure}
\caption{ Plane-wave impulse approximation version of the diagram in Fig.~1.
  }
\label{fig:Fig4}
\end{figure}

\begin{figure}
\caption{ Variation of the squared four-momentum transfer in the $({\cal E},p)$
plane for the kinematical conditions of the LSND experiment. Note the rather 
mild dependence on ${\cal E}$ and the strong dependence on $p$.
  }
\label{fig:Fig5}
\end{figure}

\begin{figure}
\caption{ The $u$-inclusive double differential cross
section for protons (solid) and neutrons (dashed) for LSND kinematics and two
different orientations of $\bbox{k}$ and $\bbox{p_N}$: (a) RFG model, (b) HM 
with harmonic oscillator wave functions and (c) as for (b), but including a 
spreading width, or equivalently, a complex nucleon self-energy. Note the 
negligible effect of the last. Note also that, although we display the results 
over the whole allowed range of $T_N$, in the LSND experiment only nucleons 
with $T_N>$ 60 MeV are actually detected.
  }
\label{fig:Fig6}
\end{figure}

\begin{figure}
\caption{ The same curves as in Fig.~6, but now for BNL kinematics. Note the 
double peak behaviour of the cross section at $\theta_N=20^0$ 
(see text for discussion). Note also that, although we display the results 
over the whole allowed range of $T_N$, in the BNL experiment only nucleons 
with $T_N>$ 200 MeV are actually detected.
  }
\label{fig:Fig7}
\end{figure}

\begin{figure}
\caption{ (a) The effect of axial-vector strangeness in the RFG and HM 
proton ejection cross sections for LSND kinematics integrated over 
angle for two values of $g_A^{(s)}$. Note the significant increase of 
the cross section induced by strangeness.
(b) The same as (a), but now for neutrons. Note the significant reduction of 
the cross section induced by strangeness. (c) The same as (a), but now for 
BNL kinematics.
  }
\label{fig:Fig8}
\end{figure}

\begin{figure}
\caption{ Ratio between the angle-integrated cross sections for proton and 
neutron ejection at LSND kinematics. Both the RFG (dashed) and HM 
(solid) results are displayed.
  }
\label{fig:Fig9}
\end{figure}

\begin{figure}
\caption{ Total cross section (integrated over angle and energy) versus the
strangeness content of the nucleon for outgoing protons (a) and neutrons
(b) at LSND kinematics and (c) for protons at BNL kinematics. Dashed lines: 
RFG; solid: HM.
  }
\label{fig:Fig10}
\end{figure}

\begin{figure}
\caption{ The ratio of the proton and neutron total cross sections for LSND 
kinematics: Dashed line: RFG; solid: HM.
  }
\label{fig:Fig11}
\end{figure}

\vfill\eject

\begin{figure}[p]
\mbox{\epsfig{file=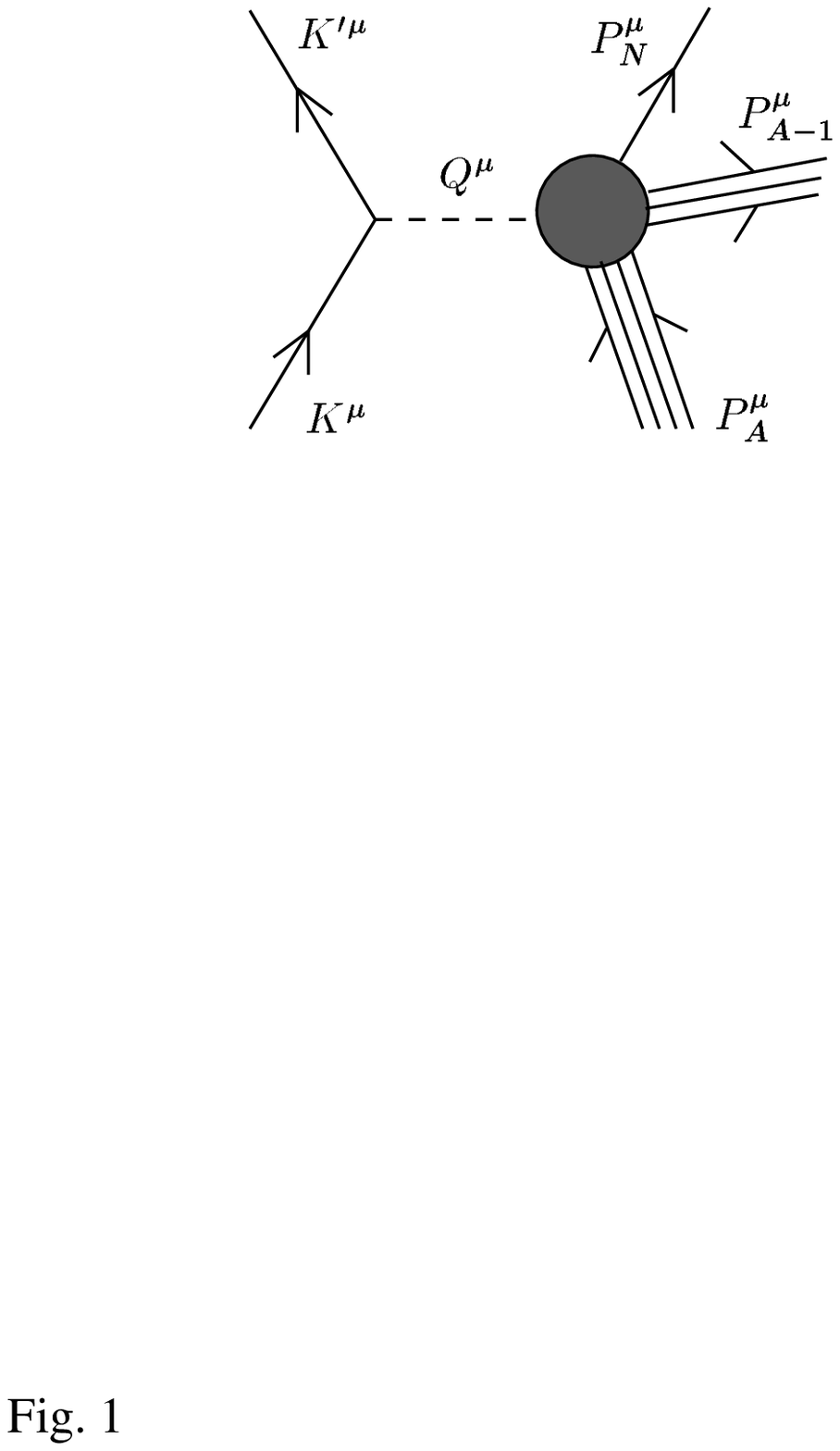,height=.9\textheight}}
\end{figure}

\begin{figure}[p]
\mbox{\epsfig{file=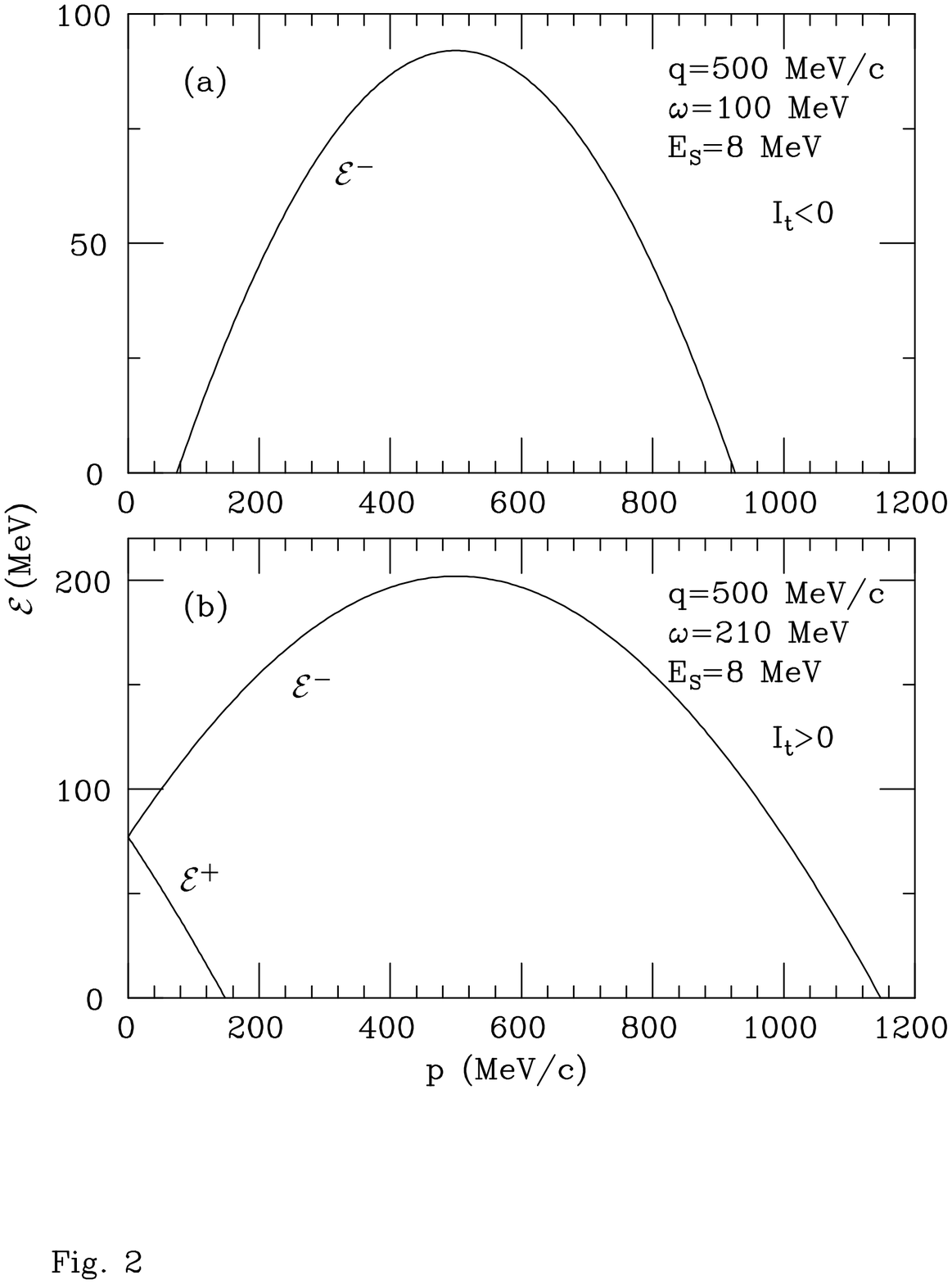,height=.8\textheight}}
\end{figure}

\begin{figure}[p]
\mbox{\epsfig{file=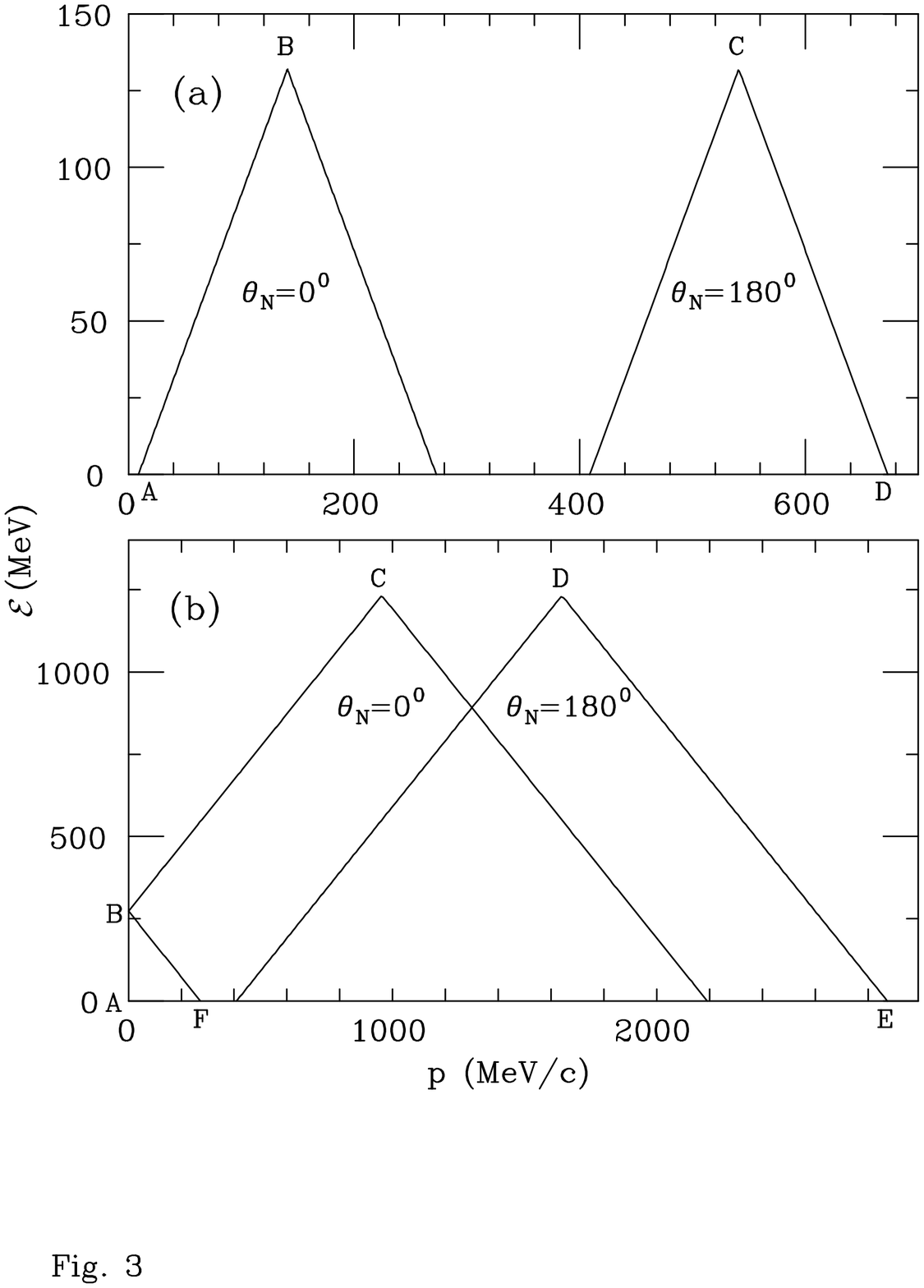,height=.8\textheight}}
\end{figure}

\begin{figure}[p]
\mbox{\epsfig{file=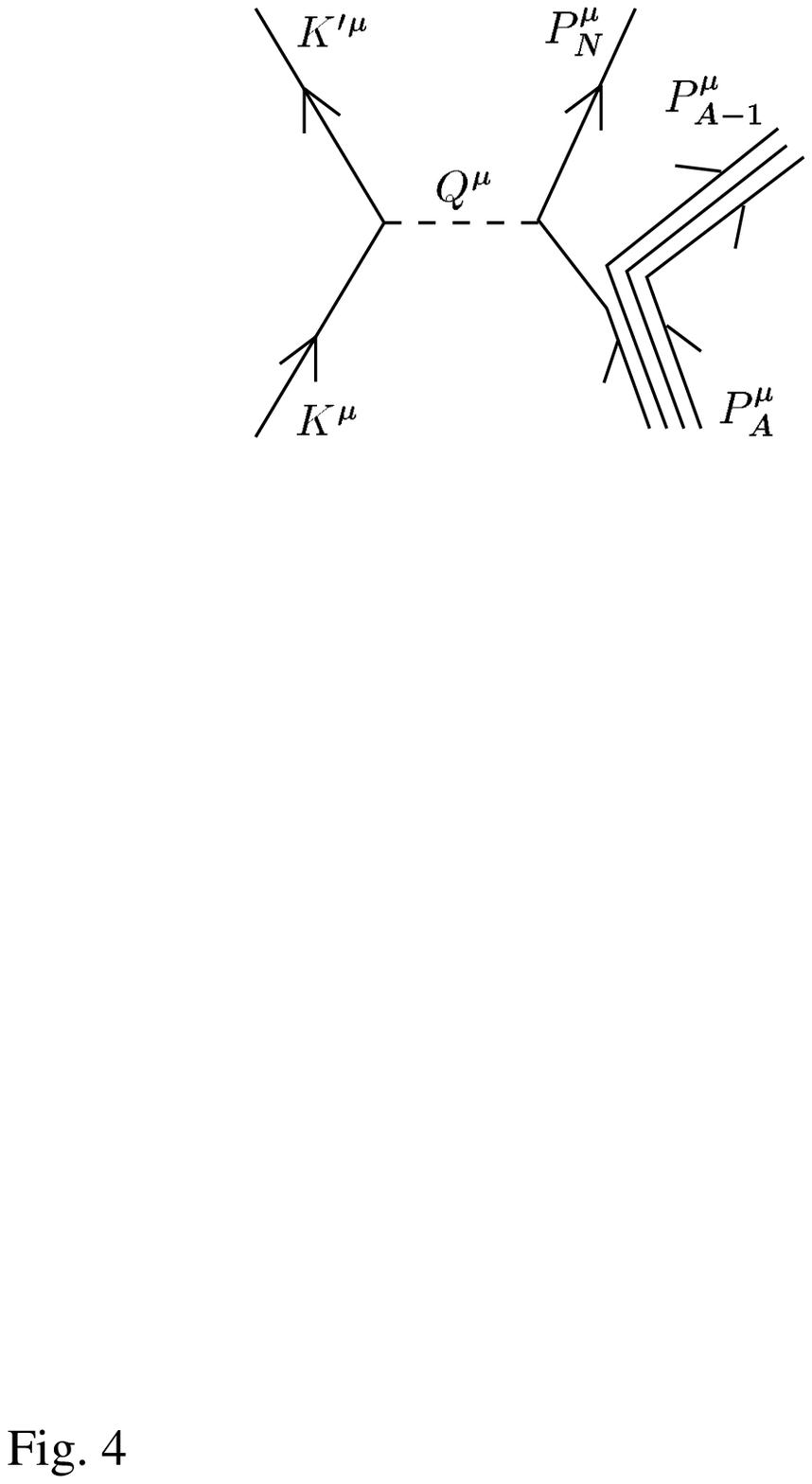,height=.9\textheight}}
\end{figure}

\begin{figure}[p]
\mbox{\epsfig{file=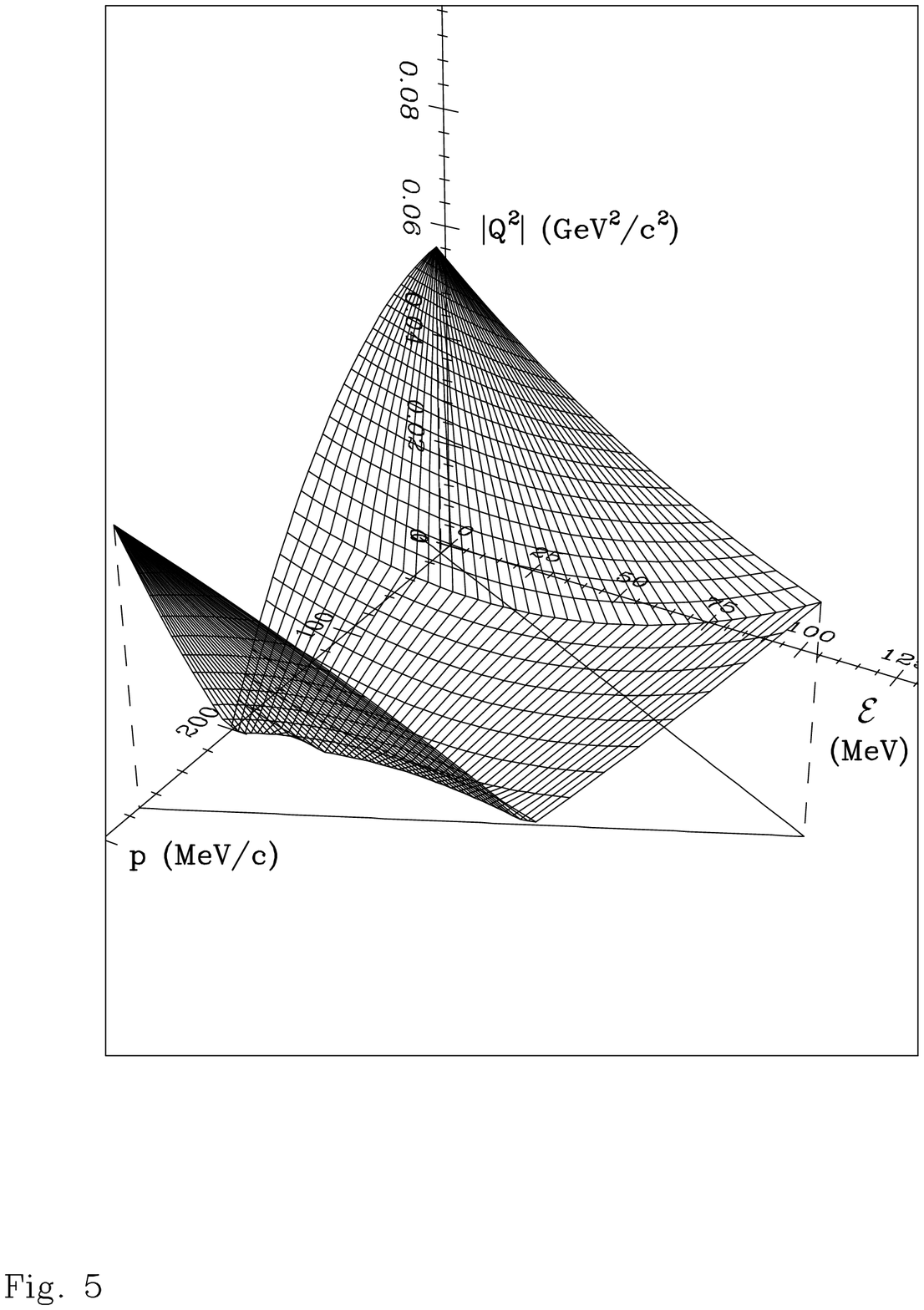,height=.8\textheight}}
\end{figure}

\begin{figure}[p]
\mbox{\epsfig{file=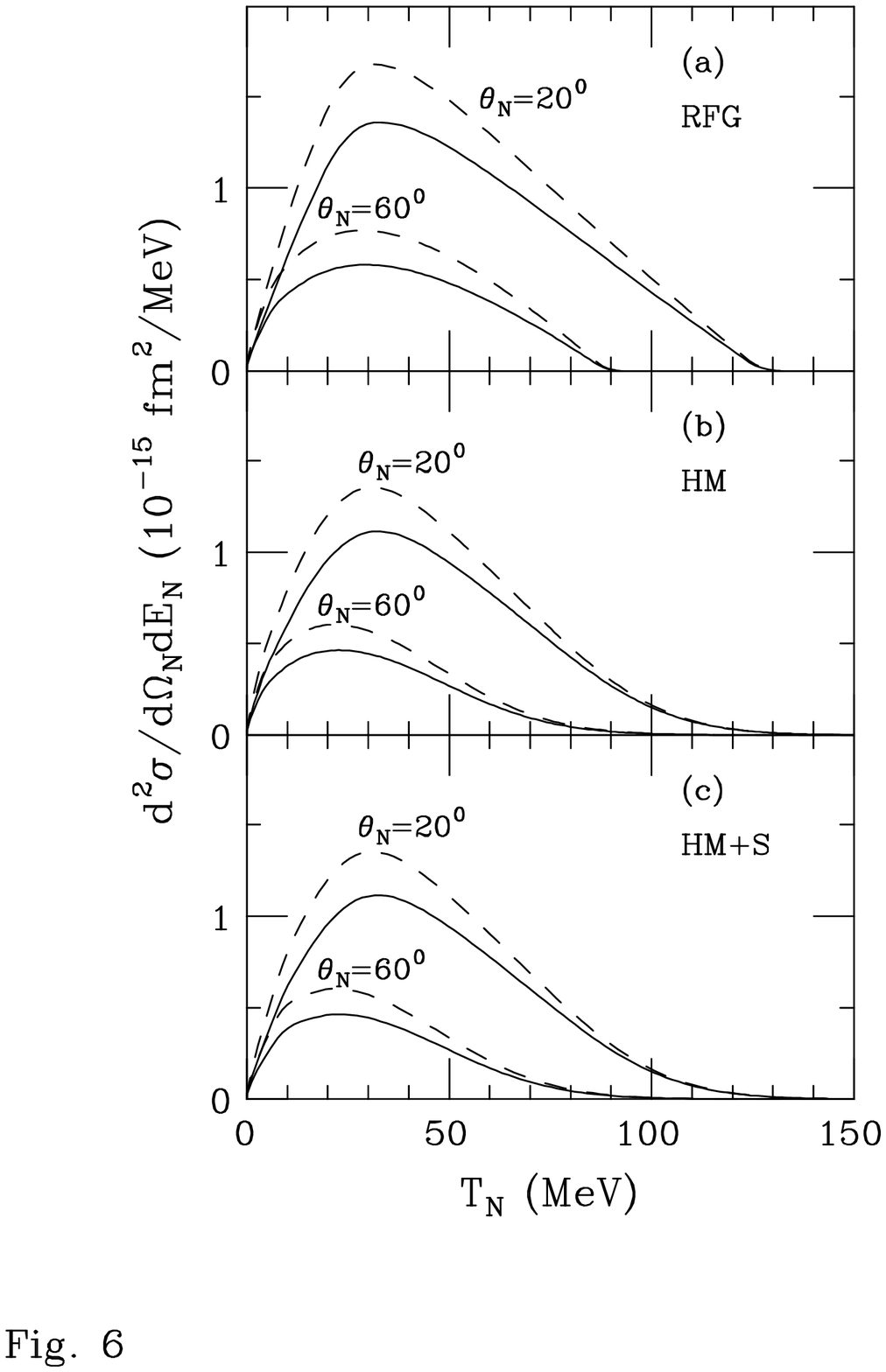,height=.8\textheight}}
\end{figure}

\begin{figure}[p]
\mbox{\epsfig{file=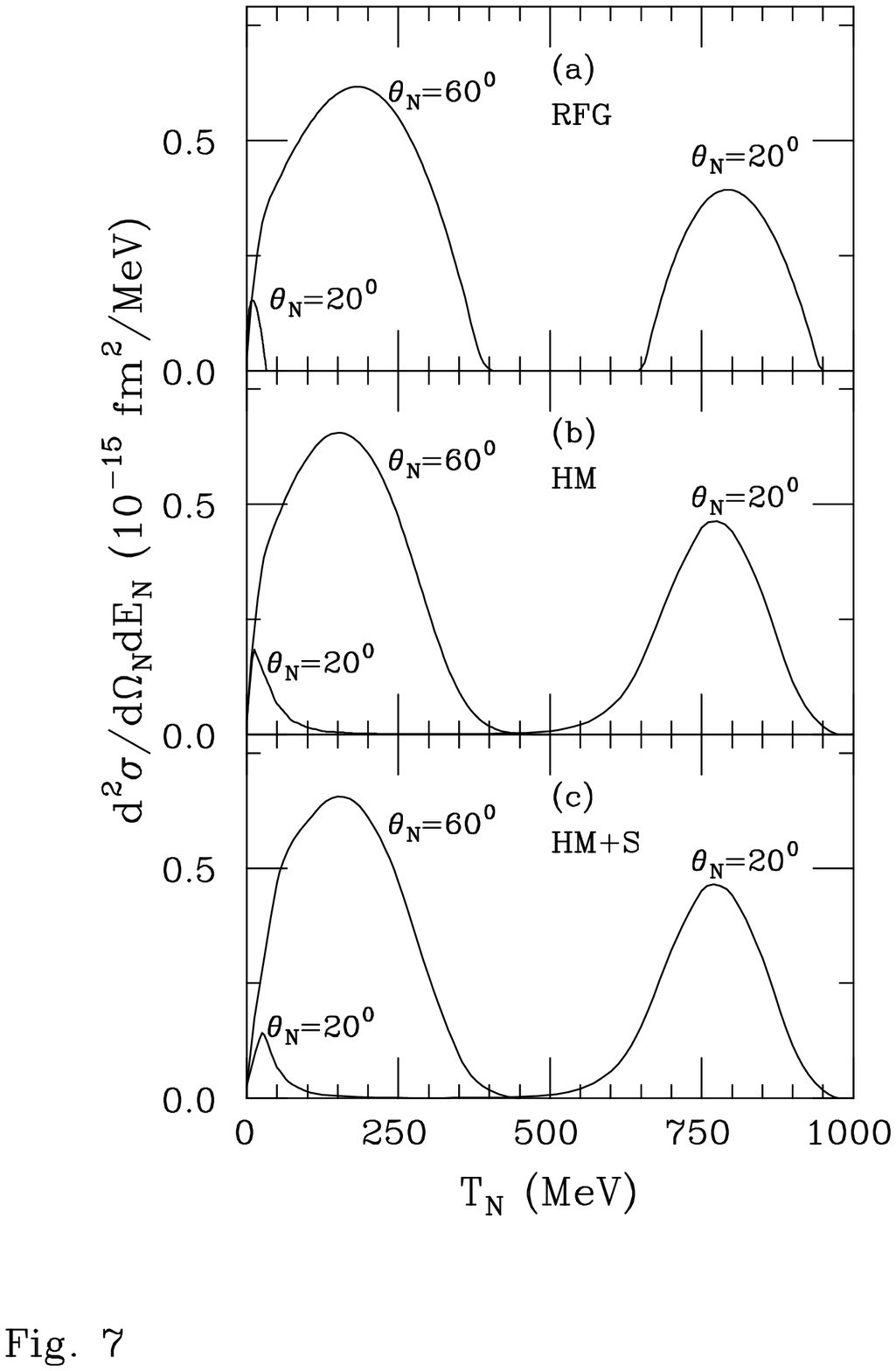,height=.8\textheight}}
\end{figure}

\begin{figure}[p]
\mbox{\epsfig{file=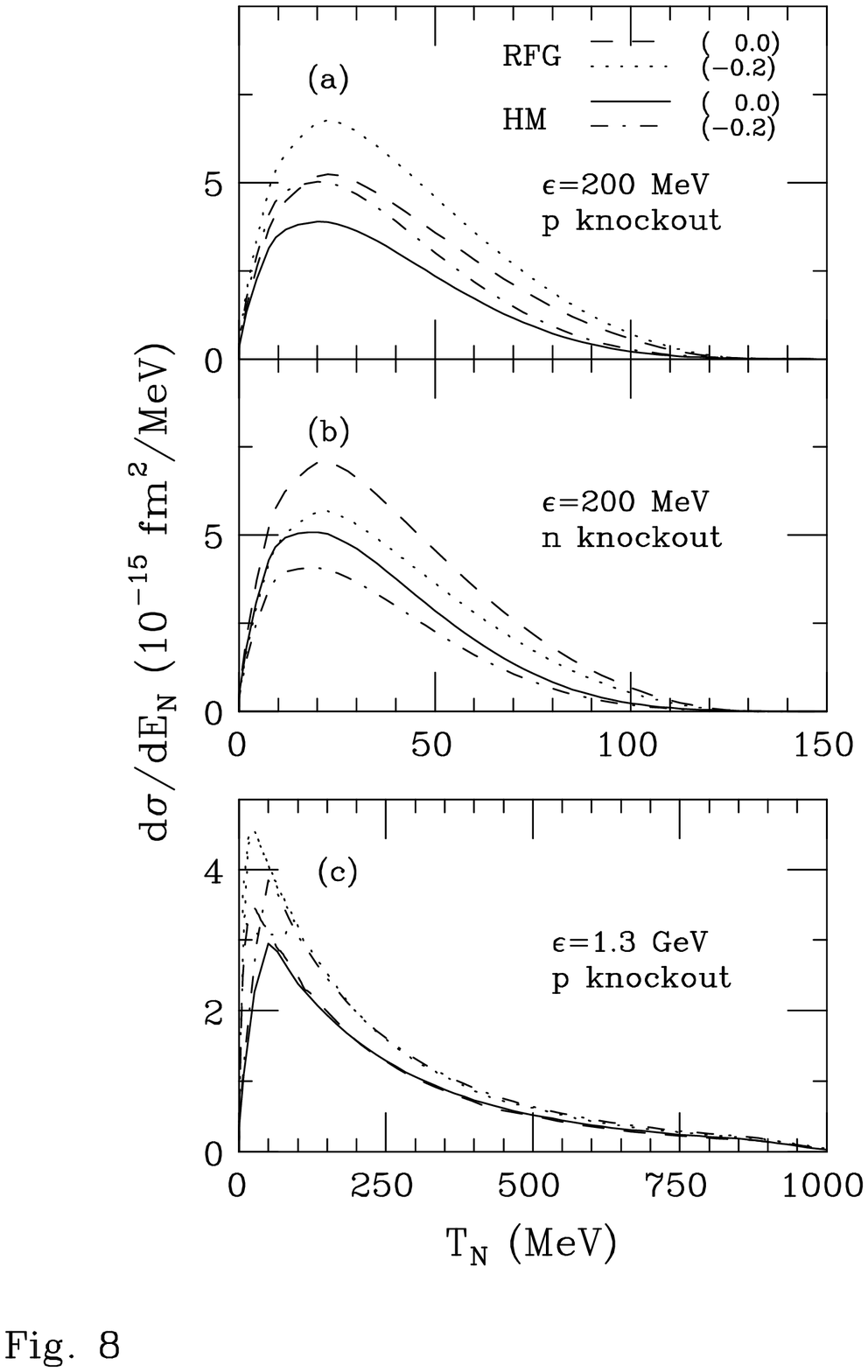,height=.8\textheight}}
\end{figure}

\begin{figure}[p]
\mbox{\epsfig{file=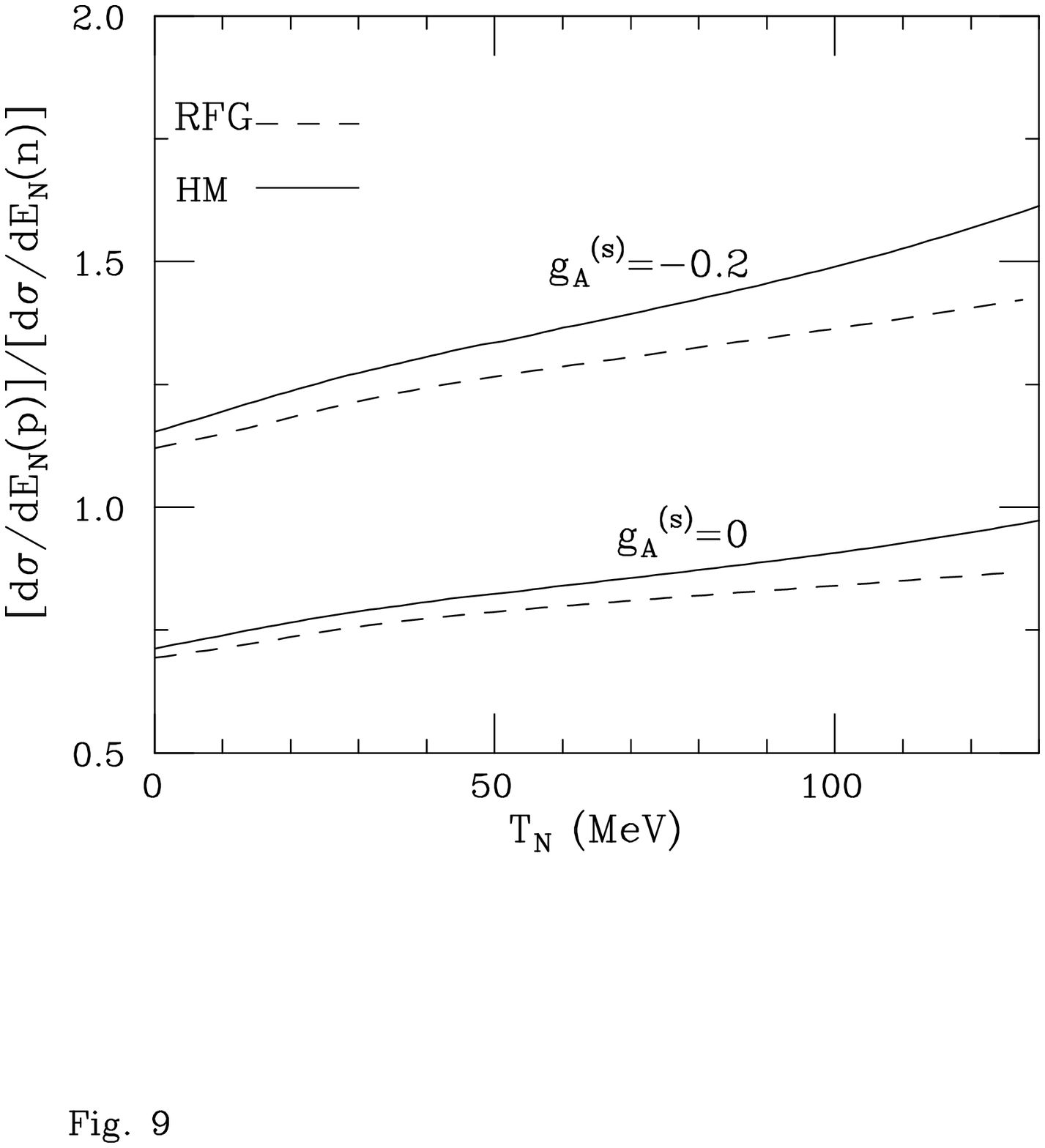,height=.8\textheight}}
\end{figure}

\begin{figure}[p]
\mbox{\epsfig{file=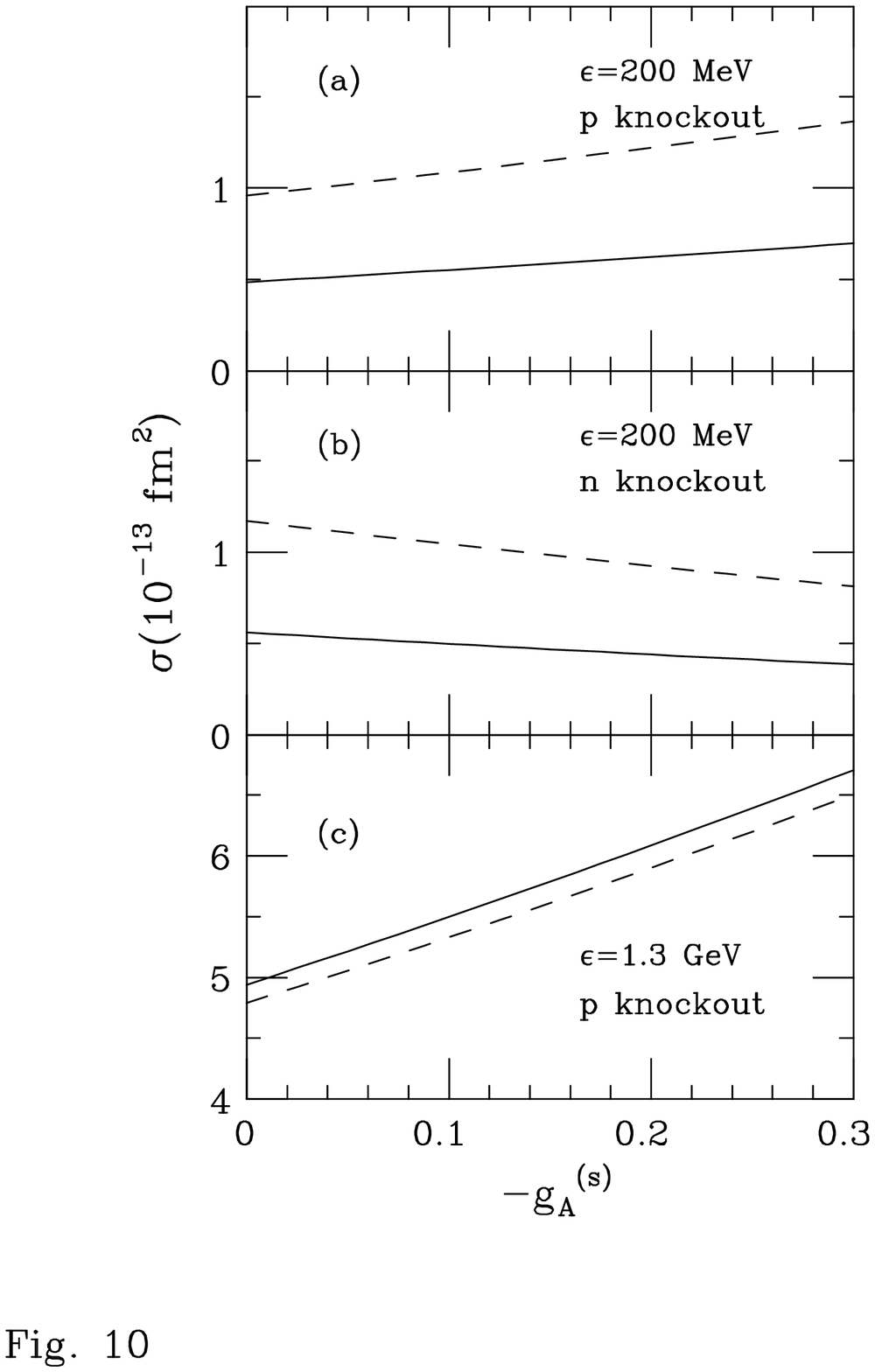,height=.8\textheight}}
\end{figure}

\begin{figure}[p]
\mbox{\epsfig{file=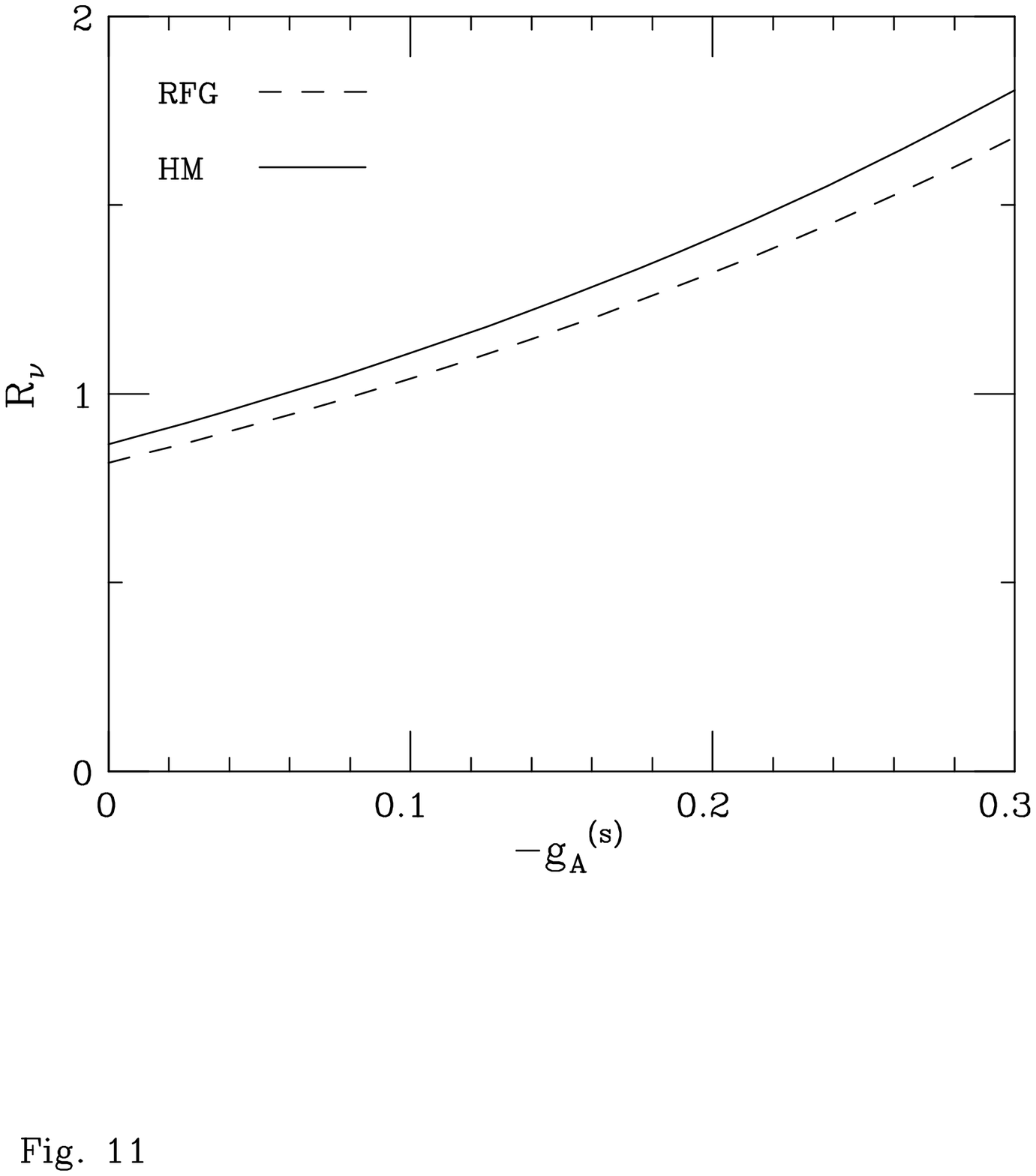,height=.8\textheight}}
\end{figure}


\begin{references}

\bibitem{Mus94}     M. J. Musolf, T. W. Donnelly, S. J. Pollock, S. Kowalski,
                    and E. J. Beise,
                    Phys.\ Rep.\ {\bf 239}, 1 (1994).

\bibitem{Ahr87}     L. A. Ahrens {\em et al}., 
                    Phys.\ Rev.\ {\bf D35}, 785 (1987).

\bibitem{Gar93a}    G. T. Garvey, W. C. Louis, and D. H. White,
                    Phys.\ Rev.\ {\bf C48}, 761 (1993).

\bibitem{Hor93}     C. J. Horowitz, H. Kim, D. P. Murdock, and S. Pollock,
                    Phys.\ Rev.\ {\bf C48}, 3078 (1993).

\bibitem{Ash89}     J. A. Ashman {\em et al}. (EMC collaboration), 
                    Nucl.\ Phys.\ {\bf B328}, 1 (1989).

\bibitem{Ade93}     B. Adeva {\em et al}. (SMC collaboration), 
                    Phys.\ Lett.\ {\bf B302}, 533 (1993).

\bibitem{Ant93}     P. L. Anthony {\em et al}. (E142 collaboration), 
                    Phys.\ Rev.\ Lett.\ {\bf 71}, 959 (1993).

\bibitem{Ada94}     B. Adams {\em et al}. (SMC collaboration), 
                    Phys.\ Lett.\ {\bf B329}, 399 (1994).

\bibitem{Abe95}     K. Abe {\em et al}. (E143 collaboration), 
                    Phys.\ Rev.\ Lett.\ {\bf 74}, 346 (1995). 

\bibitem{MAD} 	    These analyses indicate a strong correlation between
		    $\Delta s$ and $M_A$, the dipole mass parameter appearing
		    in standard parameterizations of the axial-vector form 
		    factor. In Ref. \cite{Gar93a}, a fit performed
		    forcing $\Delta s=0$ yields a value of $M_A$ higher than
		    the world average. The other three fits, in which $\Delta
		    s$ is allowed to vary, yield smaller $\chi^2$/D.O.F. and
		    values for $M_A$ consistent with the 1987 world average.

\bibitem{Lou}       LSND Collaboration, Los Alamos Proposal No. 1173, 
                    W. C. Louis, contact person (unpublished).

\bibitem{Gar92}     G. T. Garvey, S. Krewald, E. Kolbe, and K. Langanke,
                    Phys.\ Lett.\ {\bf B289}, 249 (1992).

\bibitem{Gar93b}    G. T. Garvey, E. Kolbe, K. Langanke, and S. Krewald, 
                    Phys.\ Rev.\ {\bf C48}, 1919 (1993).

\bibitem{Sam}       MIT-Bates Proposal No. 89-06, 
                    R. D. McKeown and D. H. Beck, spokespersons (unpublished).

\bibitem{Pitt}      MIT-Bates Proposal No. 94-11, 
                    M. Pitt and E. J. Beise, spokespersons  (unpublished).

\bibitem{Bei}       CEBAF Proposal No. PR-91-004, 
                    E. J. Beise, spokesperson (unpublished).

\bibitem{Bec}       CEBAF Proposal No. PR-91-017, 
                    D. H. Beck, spokesperson (unpublished).

\bibitem{Fin}       CEBAF Proposal No. PR-91-010, 
                    J. M. Finn and P. A. Souder, spokespersons (unpublished).

\bibitem{Mai}       Mainz Proposal A4/1-93, E. Heinen-Konschak {\em et al}., 
                    collaborators, D. von Harrach, spokesperson.

\bibitem{Cen96}     R. Cenni, T. W. Donnelly and A. Molinari, (to be 
                    published).

\bibitem{Jou95}     J. Jourdan,
                    Phys.\ Lett.\ {\bf B353}, 189 (1995); see also 
                    T. C. Yates, {\em et al}., 
                    Phys.\ Lett.\ {\bf B312}, 382 (1993).

\bibitem{Jen91}     E. Jenkins and A. Manohar, Phys.\ Lett.\ {\bf B255}, 558
		    (1991).

\bibitem{Dai95}     J. Dai, R. Dashen, E. Jenkins, A. V. Manohar, UCSD Preprint
		    PTH 94-19 (1995).

\bibitem{Day}       D. B. Day, J. S. McCarthy, T. W. Donnelly and I. Sick, 
                    Ann.\ Rev.\ Nucl.\ Part.\ Sci.\ {\bf 40}, 357 (1990).

\bibitem{Deforest}  T. de Forest, Nucl.\ Phys.\ {\bf A392}, 232 (1983).

\bibitem{Bjo64}     J. D. Bjorken and S. D. Drell, 
                    {\em Relativistic Quantum Mechanics} 
                    (McGraw-Hill, New York, 1964).

\bibitem{Alb88} W. M. Alberico, A. Molinari, T. W. Donnelly, E. L. Kronenberg 
  and J. W. Van Orden, Phys.\ Rev.\ {\bf C47}, 1801 (1988).

\bibitem{Mus92} M. J. Musolf and T. W. Donnelly, Nucl.\ Phys.\ {\bf A546}, 509
  (1992); 550 (1992) 564(E).

\bibitem{Mar80} W. J. Marciano and A. Sirlin, Phys.\ Rev.\ {\bf D22}, 2695,
  (1980). 

\bibitem{Mus90} M. J. Musolf and B. R. Holstein, Phys.\ Lett.\ {\bf B242}, 461
  (1990).

\bibitem{Don92} T. W. Donnelly, M. J. Musolf, W. M. Alberico, M. B. Barbaro,
  A. De Pace and A. Molinari, Nucl.\ Phys. {\bf A541}, 525 (1992).


\bibitem{Mah81}     C. Mahaux and H. Ng$\hat{\text{o}}$,
                    Phys.\ Lett.\ {\bf 100B}, 285 (1981).

\bibitem{Smi88}     R.D. Smith and J. Wambach,
                    Phys.\ Rev.\ {\bf C38}, 100 (1988).

\bibitem{Kap88}   D. Kaplan and A. Manohar, Nucl.\ Phys.\ {\bf B310}, 527
  (1988).

\bibitem{Jaf90}   R. L. Jaffe and A. Manohar, Nucl.\ Phys.\ {\bf B337}, 509
  (1990).

\bibitem{Ehr94}   B. Ehrnsperger and A. Schafer, UFTP Preprint 377-1994
		  (1994).

\bibitem{Lic95}   J. Lichtenstadt and H. Lipkin, Tel Aviv Preprint 
  	     	  TAUP-2244-95.

\bibitem{Cab}     J. A. Caballero, T. W. Donnelly and G. I. Poulis,
                  Nucl.\ Phys.\ {\bf A555}, 709 (1993).
\end{references}
\end{document}